\documentclass[12pt,preprint]{aastex}

\newcommand{\kms}{km~s$^{-1}$}
\newcommand{\subsun}{\mbox{$_{\odot}$}}
\newcommand{\teff}{$T_{\rm{eff}}$}
\newcommand{\grav}{log($g$)}
\newcommand{\etal}{{\it et al.\/}}
\newcommand{\eqw}{$W_{\lambda}$}

\newcommand{\cband}{C$_2$}
\newcommand{\ciso}{$^{12}$C/$^{13}$C}
\newcommand{\ncstars}{16}
\newcommand{\ncbahigh}{12}
\begin{document}

\title{Carbon Stars in the Hamburg/ESO Survey:  
Abundances\altaffilmark{1}}

\author{Judith G. Cohen\altaffilmark{2}, Andrew McWilliam\altaffilmark{2},  
Stephen Shectman\altaffilmark{2}, Ian Thompson\altaffilmark{2}, 
Norbert Christlieb\altaffilmark{4}, 
Jorge Melendez\altaffilmark{2}, Solange Ramirez\altaffilmark{5}, 
Amber Swensson\altaffilmark{2} \& 
Franz-Josef Zickgraf\altaffilmark{4}}

\altaffiltext{1}{Based in part on observations obtained at the
W.M. Keck Observatory, which is operated jointly by the California 
Institute of Technology, the University of California, and the
National Aeronautics and Space Administration.}

\altaffiltext{2}{Palomar Observatory, Mail Stop 105-24,
California Institute of Technology, Pasadena, Ca., 91125, 
jlc@astro.caltech.edu, aswenson@caltech.edu}

\altaffiltext{3}{Carnegie Observatories of Washington, 813 Santa
Barbara Street, Pasadena, Ca. 91101, andy, ian, shec@ociw.edu}

\altaffiltext{4}{Hamburger Sternwarte, Universit\"at
Hamburg, Gojenbergsweg 112, D-21029 Hamburg, Germany,
nchristlieb, fzickgraf@hs.uni-hamburg.de}

\altaffiltext{5}{Spitzer Science Center, Mail Stop 100-22,
California Institute of Technology, Pasadena, Ca., 91125,
solange@ipac.caltech.edu}

\begin{abstract}

We have carried out a detailed abundance analysis using
high dispersion spectra from HIRES at Keck 
for a sample of \ncstars\ carbon stars found among candidate extremely 
metal-poor (EMP) stars from the Hamburg/ESO Survey.
We find that the Fe-metallicities 
for the cooler C-stars (\teff\ $\sim$5100~K)
have been underestimated by a factor of $\sim$10 by the standard
HES survey tools.
The results presented here provided crucial supporting data 
used by \cite{cohen06apjl} to derive
the  frequency of C-stars among EMP stars.

C-enhancement in these EMP C-stars appears to be
independent of Fe-metallicity and approximately constant at
$\sim$1/5 the solar $\epsilon$(C). The C-enhancement shows 
some evidence of decreasing
with decreasing \teff\ (increasing luminosity), presumably due to mixing
and dredge-up of C-depleted material.
The mostly low \ciso\ ratios ($\sim$4)
and the high N abundances in many of these stars suggest that
material which has been through proton burning via the CN cycle 
comprises most of the stellar envelope.  

C-enhancement in this sample is associated with strong enrichment of
heavy nuclei beyond the Fe-peak for \ncbahigh\ of the \ncstars\ stars.  
The remaining
C-stars from the HES, which tend to be the most Fe-metal poor,
show no evidence for enhancement
of the heavy elements. 
Very high enhancements of lead are detected in some of the C-stars with highly
enhanced Ba.
The strong lead lines, the high Ba/Eu ratios, and the high ratios
of abundances of the diagnostic elements in the first and second $s$-process
peak demonstrate that
the $s$-process is responsible for
the enhancement of the heavy elements for the majority of
the C-stars in our sample. 

The low \ciso\ ratios and large C and N enhancements of the EMP
C-stars are more extreme than those of intrinsic AGB C-stars of near
solar Fe-metallicity, but closer to the composition of CH stars.  Our
subsample of EMP C-stars without $s$-process enhancement is
reminiscent of the R-type C-stars in the solar neighborhood; thus we
expect that they are formed by similar mechanisms.

We suggest that both the $s$-process rich and Ba-normal
C-stars result from phenomena associated with mass transfer in binary
systems.  This leads directly to the progression from C-stars to
CH stars and then to Ba stars as the Fe-metallicity increases.

\end{abstract}

\keywords{halo stars, Carbon enhancement, stars: abundances}

\section{Introduction}

We are engaged in a large scale project to find 
extremely metal-poor (henceforth EMP) stars, characterized
by [Fe/H] $\le -3.0$ dex\footnote{The 
standard nomenclature is adopted; the abundance of
element $X$ is given by $\epsilon(X) = N(X)/N(H)$ on a scale where
$N(H) = 10^{12}$ H atoms.  Then
[X/H] = log$_{10}$[N(X)/N(H)] $-$ log$_{10}$[N(X)/N(H)]\subsun, and similarly
for [X/Fe].}, by
exploiting the Hamburg/ESO Survey (HES) database.   The HES is an 
objective prism survey from which it is
possible to efficiently select a variety of interesting
stellar objects, among them EMP stars \citep{christlieb03}.
The discovery of a number
of very metal-poor, carbon-rich, objects, with diverse additional 
peculiarities, particularly
$s$-process or/and $r$-process enrichment, and the discovery of the
most iron-poor star known, HE0107$-$5240 \citep{christlieb04},
at [Fe/H]=$-$5.3, which is also very 
C-rich, recently surpassed by HE1327--2326, with similar characteristics
at [Fe/H] $\sim -5.6$ dex \citep{frebel05}, 
as well as the known C-rich binary M dwarf G77--61 established
by \cite{plez05} to have [Fe/H]
$\sim -4$ dex, all contribute
to a renewed interest in EMP carbon-rich halo stars.

Broadly speaking, when $\epsilon$(O) exceeds $\epsilon$(C) in cool
stars, 
the oxide molecules (CO, TiO, etc) dominate in the outer
layers of the stellar atmosphere.  (This is the normal
condition for solar abundance ratios.)  However, if $\epsilon$(C)
is larger than $\epsilon$(O), after the formation of CO, extra C
remains rather than extra O, and carbon compounds such as \cband, CH 
and CN will dominate.  The strong bands of \cband\
are then prominent in 
the optical spectrum of such stars, if they are cool enough,
hence the origin of the name Carbon stars (C-stars). 
Our operational definition of a C-star is one
whose spectrum shows the blue-degraded band of \cband\ at 5160~\AA,
which is the most prominent band of this molecule within the wavelength range 
of the spectra discussed here.  
If no \cband\ bands are detected, but [C/Fe] $ > 1$ dex, we denote a
star to be C-enhanced.  The strength of the \cband\ bands will be
a function of \teff,  $\epsilon$(C), and to a lesser extent,
\grav, [Fe/H] and $\epsilon$(O).

The purpose of the present paper is to carry out detailed chemical 
abundance analyses of a sample of  \ncstars\ EMP C-stars selected from the HES.
This provides a broad database to establish the Fe-metallicity for EMP C-stars.
The results presented here provided crucial supporting data 
used by \cite{cohen06apjl} to derive
the  frequency of C-stars among EMP stars.
We use the abundance ratios derived here for EMP C-stars
to discuss the origin of the C-star
phenomenon among EMP stars, which we attribute {\it{in toto}} to
phenomena associated with binary systems. 

After a description of the stellar sample in \S\ref{section_sample},
readers who are not
interested in the details of the abundance analyses should proceed to 
\S\ref{sec_individual}, then \S\ref{section_abundfe}, then skip
to \S\ref{section_ratios}. 

We favor scenarios of C-star formation among the EMP halo stars
resulting from the evolution of binary systems, including mass transfer.
The evidence supporting this is described in
\S\ref{section_ratios}.  Section \S\ref{section_disk}  
compares the derived abundances of
the EMP C-stars to those of
various types of near solar [Fe/H] disk C-stars.
The implications of our hypothesis  for C-star formation
among the EMP stars as applied to stars of higher
and lower Fe-metallicity are described in \S\ref{section_implications}.
A brief summary concludes the paper.

\section{The Stellar Sample \label{section_sample}}

The normal procedures outlined by \cite{christlieb03}
to isolate extremely metal-poor (EMP) stars from the
candidate lists produced by the HES were followed. 
In brief, candidate EMP stars were selected from the HES.
This was followed by vetting via moderate resolution spectroscopy
at 5-m class telescopes
to eliminate the numerous higher abundance interlopers.
The follow up spectra for the stars discussed here were obtained
either with the Double Spectrograph \citep{dbsp} at the Hale Telescope
at Palomar Mountain  or with the Boller and Chivens spectrograph 
on the Baade and Clay Telescopes
at the Las Campanas Observatory during the period from 2001
to the present.

These follow up spectra are used to determine an estimate of the
metallicity of the star, which is much more accurate than can be
derived from the low resolution objective prism spectra of the HES itself. 
This is accomplished via a combination of strength of absorption in 
H$\delta$ (determining \teff) and in the 
Ca~II line at 3933~\AA\ (the KP index), which determines
[Fe/H], once \teff , and hence \grav , are specified.  
The calibration of H$\delta$ plus KP index to [Fe/H] is ultimately based on
the results from high resolution abundance studies of standard stars; but
such calibrations implicitly assume that the relation between these line indices
and [Fe/H] is the same for both program and standard stars.
We denote the resulting metallicity
value as [Fe/H](HES).  The specific algorithm adopted by the HES is
described in \cite{beers99} and is identical to that
used until recently by the HK Survey of \cite{beers85} and \cite{beers92}. 
Stars were chosen for  observation
at high resolution with HIRES \citep{vogt94} at the Keck I Telescope 
primarily on the basis of low predicted metallicity; all stars
with [Fe/H](HES) $\le -2.9$ dex were put on the HIRES observing list,
as well as selected other stars of interest.

This paper is dedicated to an exploration
of these stars which turned out from their moderate resolution spectra to
be C-stars.  A more complete discussion of the selection of our
C-star sample from the HES and the frequency of C-stars within this
sample will be given in \cite{cohen06cfreq}.
We present here detailed abundance analyses  for 15 C-stars
% ***ncstars-1***  
from the HES observed at the Keck I telescope. One of these is a newly discovered
short period double lined spectroscopic binary.
We denote this group plus the dwarf C-star HE0007--1832 discussed in
\cite{cohen04} as the primary sample.  The augmented sample
also includes two C-enhanced dwarfs selected
from the HES and analyzed in the same way by our group
in our previously published papers,
HE0024--2523,
discussed in \cite{cohen02}, \cite{carretta02}, and in great detail
in \cite{lucatello03}, and HE2148--1247, discussed in \cite{cohen03},
both of which show highly enhanced lead in their spectra, plus a third
C-enhanced star whose analysis will be presented in \cite{cohen06b}.

Throughout this paper we ignore
the two known ultra-metal
poor stars HE0107--5240 \citep{christlieb04} and HE1327--2326 \citep{frebel05}.
More than 7,000 EMP
candidates were searched to turn up these two stars,
and there are no stars in the Galaxy known to us with 
$-5.2 \leq {\rm{[Fe/H]}} \leq -4.3$ dex.  
Although both of the known ultra-metal-poor stars are C-stars
with extremely large C-enhancements,  we
are not certain that they represent a continuation towards lower
Fe-metallicities of the stars discussed here, and hence we have
chosen to not consider them here.

\subsection{Stellar Parameters \label{section_params}}

In order to determine stellar parameters for these stars, 
particularly the cooler ones,
ideally we would compute a special set of model atmosphere which 
would have the abundances, particularly those of CNO, set to
values appropriate for each star.
This would ensure that
a proper accounting of the molecular absorption would be made.
We have not done this. Instead we have followed our normal
procedures described in \cite{cohen02} of matching observed broad
band photometry V--I, V--J, and V--K to predicted grids of synthetic
colors by \cite{houdashelt00}. 
\cite{cohen02} demonstrate that there is good agreement between the
Kurucz and MARCS temperature scale.  They find that the $V-K$ relations
of \cite{alonso96,alonso99} extrapolated to  EMP stars 
gives \teff\  values $\sim$100~K cooler than those adopted here for giants, 
while the deduced \teff\ from $V - K$
for stars near the main sequence turnoff are in good agreement. 

We then rely on an appropriate
12 Gyr isochrone from the grid of \cite{yi01} to obtain
the surface gravity for each star. 
The resulting stellar parameters, which have been
derived with no reference to the spectra themselves,
are given in Table~\ref{table_teff}.
By using the larger wavelength differences 
of V--I, and of V--J and V--K to determine our \teff \ values, avoiding
B--V and J--K, we achieve consistency to within $\pm150$~K between
the \teff\ determinations from each of these three colors for all
stars.  We have noticed that the B--V colors of the HES C-stars appear too red.
This behavior is expected, since 
the flux in the B band is reduced much more by molecular bands
in C-stars than
is the flux in the V band.
B--V colors thus
tend to give \teff\ that are too low,
presumably due to the effect of molecular absorption in one or both
of the filter bandpasses altering a  color which because of its
small wavelength coverage is, even under the best of circumstances, 
relatively insensitive to  \teff.
This problem with B--V colors was pointed out
by, among others, \cite{preston01}. J--K is not very sensitive to
\teff, changing in color by only 0.02 mag for $\Delta$\teff\ of 100~K.
Given that many of the HES stars are sufficiently faint that the errors
in their 2MASS photometry exceed 0.05 mag at K, we avoid the use of
J--K colors here.

The IR photometry we use is taken from 2MASS \citep{2mass1,2mass2}.  We have
obtained new photometry at V and I for
many of the stars in our sample.  We use 
ANDICAM images taken for this purpose over the past year via a service
observing queue on
the 1.3m telescope at CTIO operated
by the SMARTS consortium. ANDICAM is a dual channel camera
constructed by the Ohio State University instrument 
group\footnote{See http://www.astronomy.ohio-state.edu/ANDICAM and
http://www.astro.yale.edu/smarts.}.  Our ANDICAM program requires
photometric conditions, and additional standard star fields,
charged to our ANDICAM allocation through NOAO, are always taken for us.

Our new Andicam photometry for our sample of C-stars 
from the HES, as well
as other relevant observational data for these stars, is  
presented in Table~\ref{table_sample}.
Table~\ref{table_teff} gives the resulting stellar parameters
for these stars.

The uncertainty in \grav\ arising from our 150~K uncertainty in \teff\
depends on the slope of the relationship between \teff\ and \grav\ along
the adopted isochrone.  For stars close to the main sequence turnoff and 
for subgiants,
this is small, and the uncertainty in \grav\ is 0.1 dex.  However,
for stars along the RGB, it reaches 0.4 dex.

\section{HIRES Observations and Abundance Analysis}

Observations with HIRES at the Keck I Telescope were obtained
during several runs from Sep 2001 to June 2005.   The weather conditions
varied from night to night. 
A spectral resolution of 45,000
was achieved using a 0.86 arcsec wide slit projecting to 3 pixels in
the HIRES focal plane CCD detector.  For those stars presented here 
with $V > 15$ mag,  
a spectral resolution of 34,000 was used,
with the exception of HE1410$-$0004, which was observed at the higher
spectral resolution. 
The spectra cover the region from 3840 to 5330\,{\AA} with no gaps 
between orders
for $\lambda < 5000$~\AA, and only small gaps thereafter.
Each exposure was broken up into 1200 sec segments to expedite
removal of cosmic rays.  The goal was to achieve  
a SNR of 100 per spectral
resolution element in the continuum at
4500\,{\AA}; a few spectra have 
lower SNR.   This SNR calculation utilizes only
Poisson statistics, ignoring issues of cosmic ray removal,
night sky subtraction, flattening, etc.   The observations
were carried out with the slit length aligned to
the parallactic angle.

The recently installed 
upgraded HIRES detector designed and built by the Lick Observatory
engineering staff, led by S. Vogt, was used for 
three C-stars observed in 2005:
HE1410$-$0004, HE1443+0113, and
HE1434$-$1442.
HIRES-R was used for the first star, and HIRES-B
for the other two.  We thus obtain, 
among other desirable things, more complete spectral coverage,
reaching in a single exposure from 4020 to 7800~\AA\ with HIRES-R
and from 3200 to 5900~\AA\ with HIRES-B for the instrument configurations
we use. Note that only for one star in the present sample
does the included spectral
range reach beyond 6000~\AA.  Details  
of the HIRES exposures, including the
exposure times and the SNR per spectral
resolution element in the continuum, are
given in Table~\ref{table_sample}.

This set of HIRES data was reduced
using a combination of Figaro scripts and
the software package MAKEE\footnote{MAKEE was developed
by T.A. Barlow specifically for reduction of Keck HIRES data.  It is
freely available on the world wide web at the
Keck Observatory home page, http://www2.keck.hawaii.edu:3636/.}.
Insofar as possible, both the spectral reduction and abundance 
analyses presented here are identical to the procedures described
in our earlier paper on EMP dwarfs from the HES \citep{cohen04}.

\subsection{Equivalent Widths and Abundance Analysis \label{equiv_widths} }

The search for absorption features present in our HIRES data and the
measurement of their equivalent width (\eqw) was done automatically with
a FORTRAN code, EWDET, developed for a globular cluster project. 
Details of this code and its features are given in \citet{ramirez01}.
The strong molecular bands made it impossible to use the full spectral
range; selected regions were eliminated prior to searching for
absorption features.  This also applied to the radial velocity 
determination procedure we use, described in \cite{cohen04}.
Extensive hand checking of the \eqw\ for blending by
molecular features was necessary in many cases, such as
when the line profiles were frequently distorted by
blends, due to strong molecular blanketing, or low
S/N conditions.
The spectrum of HE1410+0213 is so severely affected by 
its very strong molecular bands that only  the region beyond
5160~\AA\ (plus a few strong lines near 4920~\AA) could be used.
For HE1443+0113 there 
is only one exposure available  which had to
be terminated at 550 sec due to deteriorating weather conditions.
It has a
very low SNR, and only the strongest features could be measured, i.e. CH,
the Na doublet, the Mg triplet, a few Fe~I lines, and two Ba~II lines.
The \eqw\ for this spectrum are more uncertain than those of the others
presented here.

The atomic data and list of unblended lines used (ignoring those in the
regions cut out due to the strong molecular bands),
are identical to those of \cite{cohen04}.
We adopt log$\epsilon$(Fe) = 7.45 dex for iron  
following the revisions in the solar photospheric abundances
suggested by \cite{asplund00}, \cite{prochaska00} and \cite{holweger01}.
Abundances were determined from equivalent widths,
except for C,
N (where we synthesized the region of the CN bandhead near 3885~\AA),
and Pb.
For C,  we synthesized the region of the CH band near 4320~\AA, which is
considerably weaker than the main bandhead of the G band
near 4305~\AA, and hence still usable even in these C-stars.
For the coolest C-stars with the strongest bands, even this
region off the main bandhead is close to saturation.
The solar abundances we adopt are those of \cite{anders89},
slightly updated as described in \cite{cohen04}.
A synthesis using our line list of CH and CN features  combined with 
the \cite{kurucz93} solar model
matches the solar FTS spectrum of \cite{wallace} with our 
initially adopted
C and N solar abundances, log$\epsilon$(C) = 8.59 dex and
log$\epsilon$(N) = 7.93 dex.  These are close to those of \cite{grevesse98},
but larger than those of \cite{asplund04,asplund05}, which are 
are 0.2 dex smaller for C and 0.13 dex smaller for N.  
Once the C and N abundances were determined for a star,
we synthesized the region of the 4057~\AA\ Pb~I line to derive the Pb abundance.

The equivalent widths and atomic parameters used
in the analysis of the primary sample of
\ncstars\ C-stars selected as EMP candidates from the HES are 
tabulated in Table~\ref{table_eqwa}, \ref{table_eqwb} and  \ref{table_eqwc}.
$W_{\lambda}$ for the additional
redder lines seen only in the three C-stars observed with the upgraded
HIRES detector are given in Table~\ref{table_eqwd}.
Occasionally, for crucial elements where no line was securely detected
in a star, we
tabulate upper limits to \eqw.

As in our previous work, we use 
the HFS components from \cite{prochaska00} for the lines we utilize 
here of
Sc~II, Mn~I, and Co~I.  For Ba~II, we adopt the HFS from \cite{mcwilliam98}.
We use the laboratory spectroscopy of \cite{lawler01a}
and \cite{lawler01b} to calculate the HFS patterns
for La~II and for Eu~II.   We adopt the isotopic and HFS shifts
for the 4057~\AA\ line of Pb~I given by \cite{vaneck03}; see her paper
for references for the laboratory and theoretical atomic physics.
\cite{mcwilliam95b} gives the HFS pattern for the NaD lines.
Although the difference between log($\epsilon$(Na)) derived
from the full HFS pattern and by just using two lines to represent
the double is small, $<$0.08 dex, we use the full HFS pattern for
these lines. 
A synthesis incorporating the list of hyperfine and isotopic components 
given by \cite{hobbs99} was used for 
the Li~I resonance line for which an upper limit for its
\eqw\  was measured in one star.  
Spectral syntheses are carried out for each of the
features with HFS to match the observed \eqw\ and 
thus derive the abundance of the relevant species.
For Pb,  because of the
strong blending by CH features, the spectral synthesis
used to determine the Pb abundance 
included lines of $^{12}$CH, $^{13}$CH, Pb and other metals.

% NaD with HFS is used for the three stars where NaD is within wavelength range
% file scr2 jlc hamburgsurvey hires summary na hfs

Recall that the amount of \cband , CH, and CN formed is dependent upon the
amount of free carbon present (i.e. the amount not locked-up in CO), and
that in general we do not have measurements
of the O abundance in these EMP C-stars.  Thus our derived C abundances
are dependent on the choice made for the O abundance through 
molecular formation and equilibrium.

The abundance analysis is carried out using a current version of the LTE
spectral synthesis program MOOG \citep{sneden73}.
We employ the grid of stellar atmospheres from \cite{kurucz93} 
without convective overshoot, when available.   We compute the
abundances of the species observed in each star using the measured
\eqw\ values and
the four stellar atmosphere
models from this grid with the closest \teff\ and log($g$) to each star's parameters.
The abundances were interpolated using results from the closest stellar model
atmospheres to the appropriate \teff\ and log($g$) for each star given
in Table~\ref{table_teff}.

\begin{figure}
\epsscale{0.9}
% Comment out the following line to embed the PS figure into the manuscript
% \plotone{/scr2/jlc/hamburg_survey/hires_sep2002/moog_files/he0012-1441_mgtriplet.ps}
\plotone{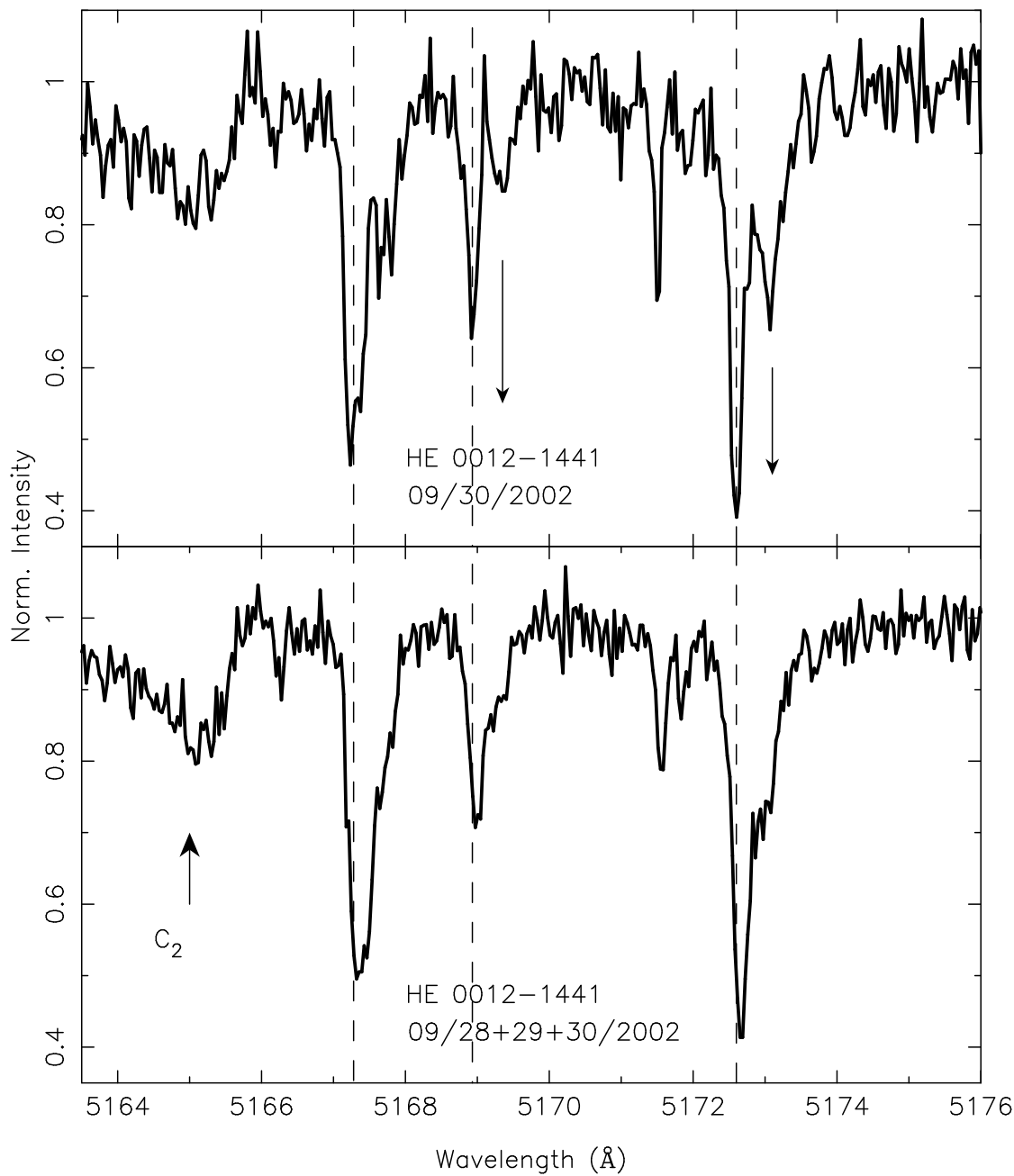}
\caption[]{Upper panel: the spectrum of the spectroscopic binary 
HE0012$-$1441 from the night of 09/30/2002 in the region of the Mg triplet.
The arrows indicate two lines from the secondary star.
lower panel: the same, but using the data summed
over 3 nights from the Sep. 2002 HIRES run.  Note the difference in the
line profiles of the Mg triplet lines (and some of the weaker lines as well).
Note also that the lines from the secondary star are noticably broader
than those from the primary.  
The dashed vertical lines guide the eye to indicate the changing
relative $v_r$ of the two components over a timespan of 48 hours.
\label{fig_binary}}
\end{figure}

Our HIRES spectra show that HE0012$-$1441 is a double lined spectroscopic binary.
Since it is rather faint, spectra were taken
on each of three consecutive nights with the intention of summing
them to reach a high SNR.   Comparison of the summed spectra for 
each of the three nights revealed the presence of double lines as well as
obvious differences in the velocity separation of the two components over
the timespan of 48 hours, as is shown in Fig.~\ref{fig_binary}.  The
$v_r$ of the primary decreased by 6~\kms\ over that timespan.
The separation of the two components was largest on the last night,
when it reached 28~\kms.  Thus, this binary  
must have a relatively short period, and is probably similar to 
HE0024--2523 (Cohen et al. 2002; Carretta et al. 2002; Lucatello et al. 2003).
Only the sum of the three 1200 sec HIRES exposures from the third night
was used to determine \eqw; this night had the largest
velocity separation, hence was the easiest from which to measure the 
\eqw\ of the primary
component.  This, of course, reduces the SNR of the spectrum below that
expected on the basis of the total integration time and below
the desired value.   We have assumed that the secondary, which contributes
perhaps 1/5 of the total \eqw\ for selected lines, does not seriously
affect the colors used to determine \teff, which may not be a valid
assumption.  Furthermore, the lines from the secondary appear to be
wider than those of the primary, suggesting a faint cool dwarf as the
secondary star.  (The secondary is too luminous to be a
white dwarf with age $\sim$10~Gyr.)  For this star only, the \eqw\ were 
not used; the \eqw\ listed in
Table~\ref{table_eqwa} for this star are for guidance only.
Instead
the abundance was determined by matching the observed line profile
for each spectral feature with the predicted one, varying $\epsilon$(X).
This ensured proper treatment of the partially blended lines due to
the second component in the binary system.  
A luminosity ratio for the two components of 4 throughout the
relevant wavelength range of the HIRES spectra was assumed 
to determine the \eqw\ for this star.

The microturbulent velocity ($v_t$) of a star can be determined 
spectroscopically by requiring the abundance to be independent of the 
strength of the lines.  However, there are fewer usable  Fe~I
lines in the complex spectra of these C-stars
than in stars with normal C and N and the same stellar
parameters due to the rejection of large regions of the spectrum where
the molecular features are strongest.  Furthermore the
uncertainties in measurement of the remaining lines are larger,
again due to possible molecular contamination and difficulties
with continuum determination that do not occur in EMP stars with
normal C and N.  Based on our as yet unpublished analyses of
a large sample of EMP giants from the HES, we  set the 
$v_t$ to 1.6 to 1.8 \kms, depending on \teff.
We checked in each case that a plot of derived Fe~I abundance as a function
of \eqw\ looked reasonable, but did not try to iterate on $v_t$ to
achieve a perfectly constant Fe~I abundance.

The abundances presented here could be improved.
Spectral syntheses could be used for additional elements.  A better
determination of \teff\ and of $v_t$ could be attempted. 
However, Table~\ref{table_slopes} demonstrates that the results
achieved here are reasonably good.  This table gives the
slope of a linear fit to the derived Fe~I abundance from each
observed line as a function of EP, \eqw/$\lambda$, and line wavelength.
Assuming a perfect analysis, these slopes should all be zero\footnote{We
ignore contributions from any issues that vary as a function of
\teff\ that may not be included in our analysis, such as non-LTE effects,
which might contribute to the measured slopes and their rms disperion.
Large contributions to the $\sigma$ of the measured slopes from 
terms such can be excluded.}.

The full
range in EP for the observed Fe~I lines is only 3 eV.
The mean slope for the derived Fe~I abundance with EP for the 11 C-stars with
entries in  Table~\ref{table_slopes} is +0.02 dex/eV
with $\sigma = 0.05$ dex/eV. We need to demonstrate that this mean and $\sigma$
are consistent with our known uncertainties. The value
of 0.05 dex/eV found for $\sigma$  corresponds to a change in \teff\ of
250~K, somewhat larger than our adopted uncertainty in \teff\ 
discussed in \S\ref{section_params} of 150~K.
However, the random component of the
uncertainty in the Fe abundance derived from a single Fe~I line
in a single star due primarily to errors  in the $gf$ value
assigned to the line is at least 0.2 dex.  This leads to a $\sigma$ for
the measured slopes Fe~I abundance versus EP
beyond that expected purely from the adopted \teff\
uncertainty.  To support this assertion we note that
the correlation coefficients for the relationship within each
star are low (${\mid}r\mid < 0.25$ in all cases). 

The slopes for the Fe~I abundance versus reduced equivalent width for
the same set of 11 stars have a mean of $-0.04$ dex with $\sigma = 0.12$ dex.
The spread in this slope is completely consistent with our adopted uncertainty
in $v_t$ of 0.2 \kms.  
The set of correlation coefficients are low (${\mid}r\mid < 0.35$ in all cases)
here also. 

The results for the abundances of typically 20 species in each star
(only 9 in HE1410+0213, and only five for HE1443+0113)
are given in 
Tables~\ref{table_abunda} to  Table~\ref{table_abundc}.
We tabulate both log$\epsilon(X)$ and [X/Fe]; our 
adopted solar abundances can be inferred directly from these tables.
The \ciso\ ratios determined from the CH and the \cband\ bands
are given in Table~\ref{table_c12c13}.

Table~\ref{table_sens} gives the changes in the deduced abundances
for small changes in \teff, \grav, $v_t$ and \eqw\
in the [Fe/H] of the model atmosphere used for an EMP giant with \teff\ $\sim$5200~K.
The last column gives 
expected random uncertainties for [X/Fe] appropriate for
for a single star, combining in quadrature 
the uncertainties in [X/Fe] resulting from
the errors in stellar parameters established in
\S\ref{section_params}, i.e. an uncertainty of 
$\pm$150~K in \teff, of $\pm$0.4 dex in \grav, of $\pm$0.5 dex
in the metallicity assumed in the model atmosphere used for the
analysis, of $\pm$0.2 \kms\ for $v_t$, and a contribution
representing the errors in the measured equivalent widths.
This last term is set at 20\% (approximately equivalent to 0.08 dex
abundance uncertainty, but depends upon line strength) for a single
detected line (which may be an underestimate for the complex
spectra of the C-stars), and is scaled based on the number of
detected lines.  The contribution of the various terms, particularly
that of \grav,, which will be smaller for hotter stars, may vary
somewhat with \teff.
Systematic uncertainties, such as might arise
from errors in the scale of the transition probabilities for an element,
are not included in the entries in Table~\ref{table_sens}.
Random errors in the $gf$ value for a particular line 
are not relevant to this calculation
provided that the same line list is used throughout.

\subsection{The \ciso\ Ratios \label{section_c13} }

We have measured the isotopic ratio \ciso\ for the C-stars from our
sample with the highest SNR spectra using the line
list for the 4300~\AA\ region of the G band of CH
as described in \cite{cohen03}. Spectral syntheses of the
features of $^{13}$CH at 4211.3, 4213.1, 4219.2,
and 4221.8~\AA\  were used. We have verified for three stars
whose HDS spectra were supplied by W. Aoki that our line
list combined with our standard analysis procedures
gives \ciso\ ratios derived from CH features differing from those
derived by 
\cite{aoki01} or \cite{aoki02a} by 15\% or less.

\begin{figure}
\epsscale{0.9}
% Comment out the following line to embed the PS figure into the manuscript
% \plotone{c2_4740_3star.ps}
\plotone{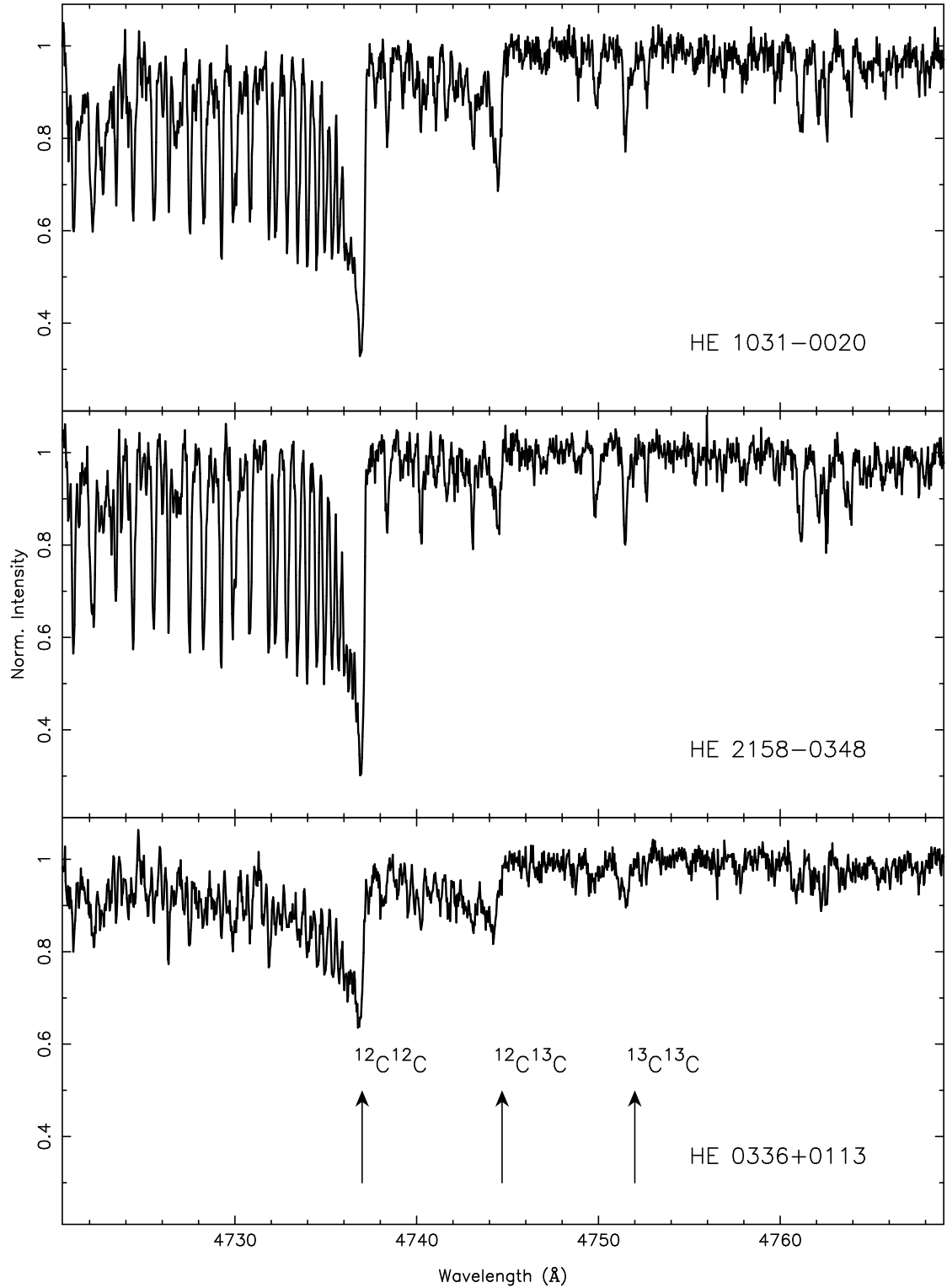}
\caption[]{The HIRES spectra of three C-stars from our sample in the region of 
4740~\AA.
The bandheads of $^{12}$C$^{12}$C, $^{12}$C$^{13}$C and $^{13}$C$^{13}$C 
are indicated.
The vertical range is the same for each panel.  The derived
\ciso\ ratio for these three stars ranges from 3.5 to 6, identical to within
the observational errors of $\pm$30\%.
\label{figure_c2iso}}
\end{figure}

Spectrum synthesis for the \cband\ bands was carried out 
based on the \cband\ line list
of \cite{querci71} and \cite{querci74}, as updated and supported
on the web site of U. J{\o}rgensen\footnote{http://stella.nbi.dk/pub/scan}.
The dissociation potential for \cband\ was taken
as 6.30 eV \citep{urdahl91}.  The isotopic line shift
depends on the ratio of the reduced mass of a diatomic molecule $AB$,
$m_A m_B/(m_A + m_B)$, for its two isotopic variants.
This ratio is 1.04 for \cband\ and only
1.007 for CH when considering $^{12}$C versus $^{13}$C.
Thus, as has been know for a long time, isotopic effects are considerably
easier to detect in certain \cband\  bands  than in those of CH.
For the G band of CH, one must study detailed profiles of individual lines 
within the band which
are often blends of multiple components of $^{12}$CH or $^{13}$CH.
The situation for \cband\ is very different. 
The strongest \cband\ band within our spectral range 
is the (0,0) Swan band at 5160~\AA, which has a very small isotopic shift.
However,
the (1,0) bandhead for  $^{12}$C$^{13}$C at 4744~\AA\
is separated
from that of $^{12}$C$^{12}$C at 4737~\AA\  by
$\sim$7~\AA, which is easily resolved even on moderate resolution spectra.
The $^{13}$C$^{13}$C bandhead is $\sim$8~\AA\ further to the red at 
4752~\AA;
it can be glimpsed in the C-stars in our sample with the smallest
\ciso\ ratios.
Plates 26 and 29 of \cite{keenan} show examples of spectra
of C-stars stars with high and low \ciso\ ratios
in this spectral region.
Figure~\ref{figure_c2iso} illustrates the ease of separating the
the bandheads $^{12}$C$^{12}$C and
$^{12}$C$^{13}$C with the present much higher resolution data. 
Any uncertainty in the band electronic oscillator
strength does not affect the determination of \ciso\ ratio.

% mu for C12C12 is 6.0, for C12C13 it is 6.24, ratio is 1.04
% mu for C12H is 0.923, for C13H it is 0.929, ratio is 1.006

Because $^{12}$C$^{13}$C is a heteronuclear molecule, and $^{13}$C has 
a non-zero nuclear spin, it has
a different number of states than does  $^{12}$C$^{12}$C, affecting
the partition function as well as the number of transitions in
a band.  
Since the spectrum synthesis program, MOOG, which we use does not 
distinguish between
isotopic molecular species it is necessary to reduce the $gf$ value of
each $^{12}$C$^{13}$C line by a factor of 2, to account for the partition
function difference with $^{12}$C$^{12}$C; $^{13}$C$^{13}$C lines would
require a factor of 4 reduction (e.g. see Amoit 1983). \footnote{Notes on
the web site of U.~J{\o}rgensen (http://stella.nbi.dk/pub/scan) suggest
that this arguement may be too simplistic, and that a factor of 4 should
be used instead to correct the $gf$ values for the $^{12}$C$^{13}$C lines.  If
true, our derived $^{13}$C abundances will need to be reduced by a factor of two.}
We note that
the band oscillator strengths for the isotopic species may not be exactly
equal, due to wavefunction differences.

We performed a sanity check on our isotopic \cband\ band line lists by 
synthesizing the $^{12}$C$^{13}$C and $^{12}$C$^{12}$C  (1,0) bandheads.
In this test we adopted a carbon abundance low enough that the bandheads were
unsaturated.  If we set \ciso\ = 1, and synthesize the spectrum
in the region of the bandhead of the (1,0) Swan  \cband\ band, 
the $^{12}$C$^{13}$C and $^{12}$C$^{12}$C bandheads 
should then
be roughly equal in strength, because although the
$^{13}$C$_2$ isotopic bandhead has twice as many lines, its partition function is a
factor two larger.  The ratio of absorption at the
appropriate resulting bandheads in the synthesized spectrum
is within
15\% of unity, as expected.

\subsection{Ionization Equilibrium and non-LTE}

\begin{figure}
\epsscale{0.9}
% Comment out the following line to embed the PS figure into the manuscript
% \plotone{fe_ion_eq.ps}
\plotone{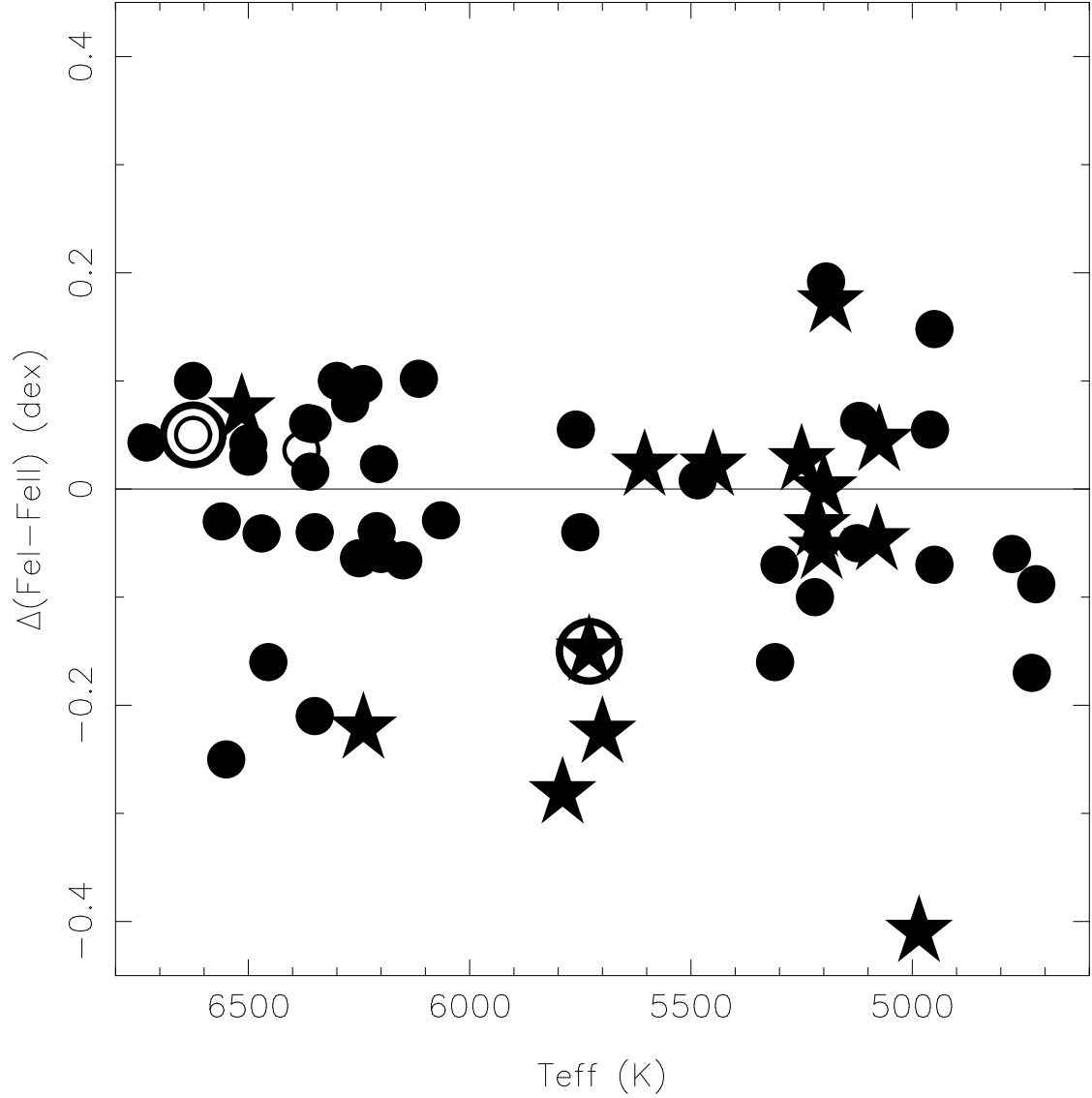}
\caption[]{The Fe ionization equilibrium for the \ncstars\ 
VMP and EMP C-stars
in the present sample.  The C-stars are indicated
as large stars, the C-enhanced stars as open circles, and the
C-normal
stars from our published and unpublished analyses as filled circles.
The three known spectroscopic binaries in this sample are circled.
(One of these is an apparently C-normal star (HE0218$-$2736) which is
too hot and too metal-poor to show
any Fe~II lines, hence does not appear in this figure.)
\label{fig_ion_eq}}
\end{figure}

Since we have not used the high resolution spectra themselves to determine
\teff\ or \grav, the ionization equilibrium is a stringent test of our analysis
and procedures, including the determination of \teff\ and of \grav,
as well as the assumption of LTE.
For the \ncstars\ candidate EMP C-stars from the HES we analyze here,
the Fe ionization equilibrium is shown in Fig.~\ref{fig_ion_eq};
we obtain a mean for
log$\epsilon$(Fe:Fe~II) $-$ log$\epsilon$(Fe:Fe~I) of $-0.07$ dex,
with a 1$\sigma$ rms scatter about the mean of 0.16 dex.
%
% Fe ion eq checked Dec 13, 2005, program fe_ion_eq.f
%
This is an extremely good ionization equilibrium for stars with
such complex spectra, and it demonstrates
the validity of our determination of stellar parameters from
photometry and isochrones.
The ionization equilibrium for Ti is almost as good, with a mean
of +0.08 dex, $\sigma = 0.28$ dex.  The dispersion falls to 0.22 dex
(and the mean becomes $-0.03$ dex) if one outlier 
with extremely weak Ti~I lines is eliminated.
% Ti ion eq checked Dec 13, 2005, program abund_range.f

The Fe abundances  derived from the
        neutral and ionized lines shift out of equilibrium by $\sim$0.25 
        dex for a 250~K change in \teff\ in this temperature regime (see
Table~\ref{table_sens}).  
Our adopted uncertainty in \teff\ is $\pm$150~K and the resulting
uncertainty in \grav\ is discussed in \S\ref{section_params}.
Table~\ref{table_sens} demonstrates
these two factors alone can give
        rise to the dispersion observed among the sample stars
in the Fe ionization equilibrium.  
        
Following \cite{cohen04}, we implement a non-LTE correction to
log$\epsilon$(Al) of
+0.60 dex  for
the lines of the Al~I doublet near 3950~\AA\ \citep{bau97}.
Only the 3961~\AA\ line can be
used for most of these C-stars; the other line of this doublet is blended 
with molecular features.  The 3905~\AA\ line of Si~I, the only suitable line
of this element in
the wavelength range covered in most of our spectra, is also heavily
blended with CH lines; we do not use it.
Si abundances have been determined only for the small number of C-stars with
the  redder wavelength coverage achieved with the new HIRES
detector, where unblended Si~I lines near 5800~\AA\
become available.  We use a non-LTE correction for Na abundances
from the 5889,5895~\AA\ doublet of $-0.20$ dex following
\cite{baumuller} and \cite{takeda}.  Only one star (HE1410$-$0004) has a spectrum
which reaches any of the O~I features, yielding an upper limit to
the 6300~\AA\ forbidden line and a marginal detection of the strongest
line in the triplet at 7772~\AA.  \cite{kisselman01} pointed out
the need for non-LTE corrections for the IR triplet line, and we use
those calculated by \cite{takeda03}.  We adopt a non-LTE correction for O
for this star of $-0.2$ dex.

% correction delta = a 10**(b x eqw(mA)), table 3, a=-0.06, b=0.068,
% about -0.2 dex.

\subsection{Comparison with Previous High Dispersion Analyses}

Two of the \ncstars\ C-stars studied here are rediscoveries of stars found
in the HK Survey (Beers, Preston \& Shectman 1985, 1992), 
and have been previously observed at high dispersion.
HE0058--0244 (CS 22183--015) was analyzed by \cite{johnson02}.
They relied on
stellar parameters determined from the spectra themselves; their
adopted \teff\ (5200$\pm$100~K) is 400~K lower than our value, and their \grav\ 
is correspondingly 1 dex
higher to preserve ionization equilibrium.  The difference in their
derived [Fe/H], which is 0.35 dex lower than our value, is 
due entirely to the
differences in the adopted stellar parameters. We attempt to compare
[X/Fe], modifying their values to our adopted \teff,\grav\
using the sensitivity table (Table~\ref{table_sens}).  With these corrections,
which in some cases are large, we find pretty
good agreement (within 0.25 dex), except for  
[Y/Fe] and [La/Fe], where our abundances are 0.4 dex lower than theirs.

\cite{norris97}, \cite{bonifacio98}, \cite{preston01}
and recently \cite{aoki02b} observed 
HE2356--0410 (CS 22957--027).
The first two groups used B--V to establish \teff; 
they both use a value 350~K
cooler than that we adopted here. \cite{preston01} uses a hybrid
method with B--V corrected for molecular absorption
to determine \teff, while \cite{aoki02b} used
B--V and V--K for this purpose, ending up with a value for \teff\ only
100~K lower than ours.  These differences in adopted
stellar parameters directly produce the differences
in derived metallicity: the first two analyses yield
[Fe/H] values 0.3 dex lower than adopted here, while that of the
last is only 0.05 dex lower.  Although there is overall good
agreement for the C abundances, the derived [N/Fe] ranges
over 0.9 dex among the five analyses.

We compare our derived abundance ratios with those of \cite{aoki02b}, as the
stellar parameters adopted in these two analyses  are similar.
Their \ciso\ ratio is 8$\pm$2, in reasonable agreement with our value
of 4.0$\pm$1.3 (see
Table~\ref{table_c12c13}).
Our [C/Fe] is 0.2 dex lower than theirs, while our derived [N/Fe] is 
a similar amount larger, as it must be to compensate in order to fit the
CN band strength.  The abundance ratios for all
the  species in common agree fairly well,
with [Al/Fe], [Ca/Fe], [Sr/Fe] and [Ba/Fe] showing the largest
differences, $-0.42$\footnote{The difference in adopted non-LTE
correction for the lines of the 3950~\AA\ doublet of Al~I has been 
removed.}, $+0.58, -0.42$ and +0.45 dex respectively
for the values of [X/Fe] derived here minus those of \cite{aoki02b}.
Large differences also occur in abundance ratios for many species
in comparing our results with the other
earlier analyses.

To isolate the cause of the large differences between the various
analyses of these C-stars, we have compared our measured \eqw\ with those
published, when available.  For HE0058--0244, 
the measured \eqw\ for the 12 weak lines in common  (mean \eqw\ of 27.7~m\AA)
tabulated by \cite{johnson02} for $n$-capture elements
agree with ours with
a mean difference of 1.1~m\AA\ and a $\sigma$ of 3.2~m\AA.
The only strong line in common is the 4554~\AA\ line of Ba~II with
measured a \eqw\ of 177.5 and 166.2~~m\AA\ in the two studies.
For HE2356--0410 we have 18 lines in common with those tabulated by
\cite{norris97}.  The \eqw\ again agree well, with
a mean difference of 0.3~m\AA\ and a $\sigma$ of 12.2~m\AA.
(The set of lines in common in this case are in general stronger lines, 
with a mean \eqw\ of 62~m\AA.)
The agreement with the \eqw\ for this star tabulated by 
\cite{bonifacio98} is also very good, with $\sigma$ of 6.1~m\AA.

Thus the differences in deduced abundances between the analysis presented
here and those previously published for these two C-stars are not
due to differences in measured \eqw.
They must arise from the choices made
for the stellar parameters and in the details of the
abundance analyses.  
In spite of these discrepancies, the
overall characteristics of the abundance distribution in these two
C-stars
are inferred as identical by each of the analyses.
All five groups, for example,
agree that HE2356--0410 has an extremely large enhancement of C, and
has a very low [Ba/Fe].  The
deviations from ``normal''
EMP stars are in general and for this particular star very large,
larger than the errors made by any of the independent
analyses.  

% My Al non-LTE correction of -0.60 dex is already taken into account
% in the comparison for HE2356-0410

We previously published an analysis of the dwarf C-star HE0143--0441
in \cite{cohen04}.  The analysis presented here supersedes that one;
the adopted \teff\ is 130~K cooler due to acquisition of better optical
photometry in the interim  and the \eqw\ have also been rechecked
for molecular blends since our earlier effort. 
The resulting 
[Fe/H] is 0.14 dex smaller than that of our previous work.  
The abundance ratios [X/Fe] derived from
our two analyses are in good agreement, except for N. 
It appears there was a typo in the entry for
log$\epsilon$(N) in Table~5 of \cite{cohen04} which
is corrected in Table~\ref{table_abunda} here.

S. Lucatello's PhD thesis
\citep{lucatello06} will present a detailed abundance analysis for
five of the C-stars analyzed here.  That analysis
should be definitive, with extensive use of spectral syntheses
and maximum care in all aspects. The Si abundance should be recoverable
with such syntheses, and a careful synthesis of the region of the 3961~\AA\ line
of Al~I would improve the Al abundances presented here.

\section{Comments on Individual Elements \label{sec_individual}}

\subsection{Iron \label{section_abundfe} }

We confirm that our
[Fe/H] determinations are largely free of molecular contamination by looking
at the derived Fe-abundance in regions where molecular bands are absent
as compared to those where they are (weakly) present.  Regions where
the molecular bands are strong in the spectrum of a sample star were
ignored.
Every star in our sample was checked to make sure that the Fe~I abundance
deduced from lines redward of 5160~\AA\ to the end of our spectral
coverage, a region within which there are no
molecular features, was the same as that for lines to the blue.
For only two stars did a possible systematic difference appear,
and it was only 0.1 dex,
with the redder lines giving slightly lower Fe abundances.  
This supports the validity of our Fe-abundances.

\begin{figure}
\epsscale{0.9}
% Comment out the following line to embed the PS figure into the manuscript
% \plotone{/scr2/jlc/hamburg_survey/paper_cstars_2005/apjl_fig1_v2.ps}
\plotone{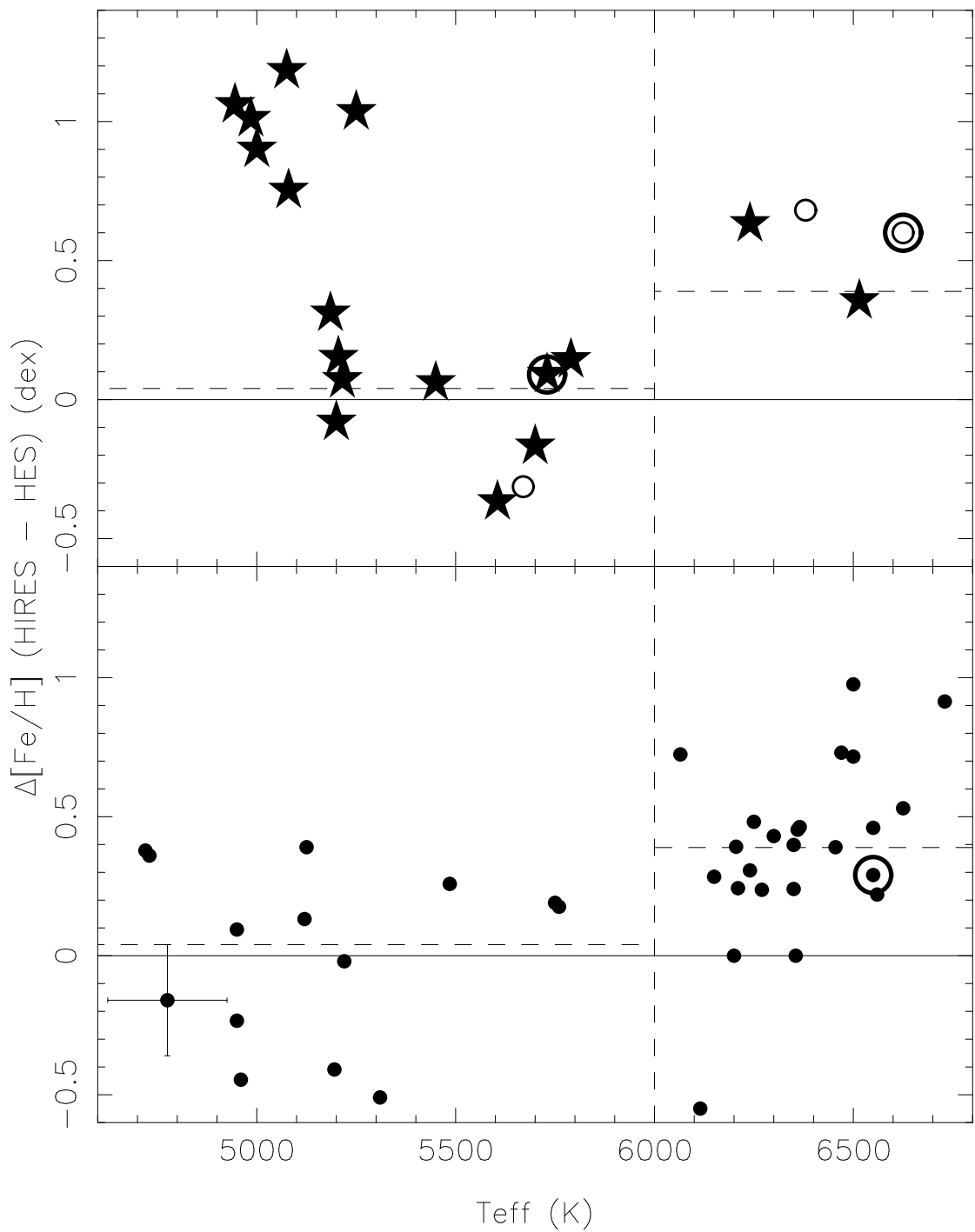}
\caption[]{The difference between [Fe/H](HES) and [Fe/H](HIRES) is shown
as a function of \teff\ for the C-stars (upper panel)
and for the C-normal
stars (lower panel) for those EMP candidates from the HES with analyses based
on Keck/HIRES spectra. The symbols are those of Fig.~\ref{fig_ion_eq}.
The vertical dashed line separates the giants
from the dwarfs, while  the horizontal dashed lines are represent
the mean $\Delta$ for the C-normal giants and for the C-normal
dwarfs.  A typical error is indicated for a single star.
\label{fig_delta_feh}}
\end{figure}

The [Fe/H] values derived here are in some cases considerably
higher than those predicted by the algorithm used on the
moderate resolution HES follow-up spectra.  Fig.~\ref{fig_delta_feh}
shows $\Delta$[Fe/H], the difference between
[Fe/H] as determined from a detailed abundance analysis of
high dispersion spectra  versus that from the application of
the \cite{beers99} algorithm to the moderate resolution spectra.
Initially, both for the HES and for the HK Survey, the B$-$V color
was used to indicate \teff.  Such a procedure is very convenient
for the HES, for example, as rough colors can be measured directly
from the objective prism spectra. This procedure, however,
is a disaster for C-stars, as the B
bandpass is much more affected by molecular absorption from CH and CN,
than is the V bandpass.  Spuriously red B$-$V colors lead to spuriously
low deduced \teff,  which in turn lead to spuriously low deduced
Fe-abundances. In practice
this affects all C-stars cooler than 6000~K, and almost certainly some
even hotter than that.  The literature is full of references
to C-star abundance analyses where the resulting high resolution
[Fe/H] grossly (by $\sim$1 dex) exceeds [Fe/H](HK), see, for example,
\cite{norris97} or \cite{hill00}, for which the relevant [Fe/H](HK)
are given in \cite{barbuy97}.  
The origin of this problem was realized several years ago
by several groups, see, for
example, \cite{preston01}.
Both the HES and the HK Survey then switched to using
the strength of absorption at H$\delta$ as a \teff\ indicator.

However, there is still a problem for the cooler C-stars, as is
shown in  Fig.~\ref{fig_delta_feh}.  The five
C-stars HE0212$-$0557, HE1031$-$0020, HE1434$-$1442,
HE1443+0113 and HE1509$-$0806 
show  $\Delta$[Fe/H] $\sim$1 dex.
Something is still wrong, but now
only the cooler giants, \teff\ $\sim 5100$~K,
and not all of them, are affected.  As was shown by 
\cite{cohen06apjl},
the problem is the molecular absorption in the specific bandpasses
used, particularly in the red continuum bandpass for the H$\delta$ index.
For the most extreme C-stars in our sample, the HP2 index 
measuring the
H$\delta$ absorption defined and used by the HK Survey
becomes negative (i.e. implies that H$\delta$ is in emission),
which is not the case when one examines high dispersion spectra.
This again leads to spuriously low \teff\ estimates and hence to spuriously
low deduced Fe-abundances from the moderate resolution follow up HES or HK
Survey spectra.
Both CN and CH contribute to the absorption there, with that
of CN dominating at solar metallicity in the relevant \teff\ range.  
At the low Fe-metallicities
considered here, the relative contribution of CN and of CH will depend
primarily on \teff, with C/N ratio and Fe-abundance also playing a role.
The five C-stars with large $\Delta$[Fe/H]
(those discrepant in Fig.~\ref{fig_delta_feh}) are the five stars
with the strongest absorption over the specific spectral
region of interest (4144 to 4164~\AA).

In the regime of KP and HP2 corresponding to EMP giants, 
a change in HP2 of 0.5~\AA\ can
produce a change in predicted [Fe/H](HES) of 0.5 dex.
The filter bandpass of HP2 is 12~\AA\ wide \citep{beers99}.  Thus, a 0.5~\AA\
error in the measured HP2 index corresponds to a 
4\% error in the continuum level. Looking at the spectra of the coolest
C-stars in the 4000-4200 A region shown in Fig.~2 of \cite{cohen06apjl} in 
the relevant region for the
feature and sideband bandpasses of HP2, it is difficult to see how
an underestimate of the continuum level of this size will not occur.

Thus, the algorithm adopted by the HES, and until recently the HK Survey
(Rossi et al. 2005), to deduce a Fe-metallicity from the low dispersion spectra,
systematically underestimates [Fe/H](HES) by a factor of $\sim$10,
for certain cool C-stars (\teff\ $~\lesssim$5100~K). 
The important implications of this are: the overestimate of the frequency of
C-stars among EMP stars, and the overestimate of the yield of EMP stars in the
HES and, by implication, the HK Survey.  These issues are discussed briefly in
\cite{cohen06apjl} and will be discussed at length in \cite{cohen06cfreq}.  

We demonstrate that the systematic [Fe/H] underestimate for EMP C-stars does not
arise from the random uncertainty in the measurement of the HP2 indices.
Comparison of HP2 indices measured from moderate dispersion spectra
for 57  stars, most of which are C-normal, with observations on different 
runs at  the P200 or observed at both the P200 and Magellan telescopes show a 
mean difference in measured HP2 indices of $0.18$~\AA, with a rms dispersion 
about the mean of 0.65~\AA; details will be presented in \cite{cohen06cfreq}.
Also note that the moderate resolution spectra
of the 5 C-stars which show large $\Delta$[Fe/H](HES)
are from four different runs with the
Double Spectrograph on the Hale Telescope.
% Date of P200 spectra
% HE0212-0557  Oct 2001
% HE1031-0020  March2002 
% HE1434-1442  April 2004
% HE1443+0113  April 2004
% HE1509-0806  May 2002   

The slight overlap of high and low
$\Delta$[Fe/H](HES) values at the boundary in \teff\ where
this effect becomes important ($\sim$5200 K) 
can be explained as resulting from observational 
uncertainties; recall that our adopted uncertainty in \teff\
for these C-stars is 150~K.  Furthermore, this effect depends on the C abundance, 
the C/N ratio, and to a smaller extent [Fe/H], although the primary
dependence is on \teff. 
Clearly, while using H$\delta$ is better than using B--V as a \teff\ indicator,
it has its limitations, particularly for cool C-rich giants, as shown here.
Using a V--K color is better.  J--K is not useful for the faint stars found 
in the HES; the errors of the 2MASS database are too large compared to the 
sensitivity of J--K to \teff\, which is, as discussed
in \S\ref{section_params}, small.  This statement may not hold for the 
HK Survey, where the stars are in the mean significantly brighter 
than the HES, and hence the 2MASS errors are much smaller.
 
%
% (HE1434-1442) 
% has an uncertainty in Teff higher than typical (no photometric V mag yet).
% Andicam VI expected soon
%

\subsection{C and N}

Since a band of CN is used to derive the N abundance, 
the N abundance is linked to the choice of C.
Systematic errors not included in Table~\ref{table_sens} in
the C and N abundances are possible in the case of unusually 
large oxygen abundance,
because the CN and CH densities depend upon the amount of free carbon
left given CO formation.

HE1150--0428 has extremely strong CN bands; the bandheads at 3885, 3875
and 3865~\AA\ are all present and the first two of these 
reach maximum absorption of $\sim$85\% of the continuum.   The
continuum was very hard to define in this region of the spectrum of this
star.  Combining that with saturation issues, the N abundance
this star is not well determined; appropriate errors might be
log[$\epsilon$(N)] = 7.15~(+0.5,$-0.3$) dex.

The determination of \ciso\ from the \cband\ and CH  bands is
described in \S\ref{section_c13}.
Fig.~\ref{fig_c2band} displays the measured absorption of the
stellar continuum at
$^{12}$C$^{12}$C and $^{12}$C$^{13}$C bandheads for the C-stars
in our sample;  see also
Figure~\ref{figure_c2iso}.
The deduced isotopic ratios for C derived
from these two bands 
are shown as a function of \teff\ in Fig.~\ref{fig_c12c13}.
\ciso\ is easy to determine
from the 4740~\AA\ \cband\ band, and the many values given in 
Table~\ref{table_c12c13} demonstrate that the \ciso\ ratio is low,
with a typical value of 4.  The isotopic ratios 
for our sample of C-stars as determined from \cband\ bands, ignoring
the lower limits, are consistent to within 1.5$\sigma$ with
a constant value of \ciso\ of $\sim$3.5.  Similar values have been
found among luminous moderately metal poor field giant stars
with normal C-abundances and with luminosities near the tip
of the RGB by \cite{carretta00}.
The hottest stars in the sample
yield only lower limits to \ciso\ using either of the molecular
features.  More measurements of \ciso\ ratios at the extremes of
the range of \teff\ would be required
to search for any trend with \teff.

\subsection{Barium}

In many cases, the Ba~II lines are very strong, and the resulting
derived Ba abundances must be regarded as quite uncertain.  Their
HFS corrections are sometimes large and vary considerably
with \eqw.  The
HFS corrections calculated by \cite{mcwilliam98} which we adopt
are for a $r$-process
isotopic distribution.  We have rescaled them for the $s$-process
Ba isotopic distribution; this in general reduces the
deduced Ba abundance by $\sim$0.1 dex.

\subsection{Lead \label{section_lead}}

There is only one usable Pb~I line in the spectral region we cover.  This line,
at 4057.8~\AA, is badly blended by CH features in these C-stars. 
Our spectral synthesis for this feature uses
the isotopic and HFS pattern for Pb described in 
\S~\ref{equiv_widths}, as well as features of
$^{12}$CH, $^{13}$CH and various atomic species.
A reasonable uncertainty for our Pb measurements based on
spectral synthesis is $\pm0.3$ dex.  Non-detections, in cases where
there is no problem indicated by notes in Table~\ref{table_abunda} to
\ref{table_abundd}, correspond to upper limits of log[$\epsilon$(Pb)] = +1.5 dex.

\subsection{Use of Strong Lines}

It is desirable in carrying out a detailed abundance analysis to use
only absorption lines with \eqw\ less than $\sim$170~m\AA\ 
to keep the errors as small as possible.  Stronger lines will be formed in the
outermost layers of the stellar atmosphere, where the 
$T(\tau)$ relationship is more uncertain, and where LTE is less likely
to prevail. Hence the \eqw\
predicted from a model atmosphere for such strong lines are more uncertain, 
as is the derived
abundance of the species from which the line originates.  However,
the wavelength coverage of our spectra, almost all of
which were taken prior to the HIRES detector upgrade, is restricted,
and CH, CN and \cband\ molecular bands in the spectra of these stars further
cut down the useful wavelength range.  Some elements have very few 
detectable lines of any state of ionization
within the allowed region.  In a few cases only strong lines are
available, while in others one or two weak lines are sometimes present together
with the strong ones, at least for a few of the C-stars in our sample.

Examination of Tables~\ref{table_eqwa} to \ref{table_eqwd} reveals the
elements of concern.
The Na~I D lines are too strong for reliable abundance analysis
in the spectrum of our coolest C-star,
HE1443+0113, and are the only lines detected of that species in the only
available HIRES spectrum of that star, which has low SNR.
Two lines of the Mg triplet at 5170~\AA\ are
always detected (the third is blended and not used) and are sometimes stronger 
than 170~m\AA, 
but often one or more of the weaker subordinate Mg~I lines
are seen as well. 
The Sr~II line at 4077~\AA\ is the only one measured in many 
of the sample stars, as the 4215~\AA\
line is often swamped by CN.  In the most $s$-process rich cool C-stars, this line
exceeds the \eqw\ cutoff suggested above; unfortunately there are no
other detectable Sr lines in the available wavelength region.
The  Ba~II lines at 4554 and 4934~\AA\ are extremely strong, far beyond
the limit in \eqw\ suggested above, in several of the cooler Ba-rich
C-stars. But in many of these, the weaker 4130~\AA\ line is seen as well,
and in the one star with a HIRES-R spectrum, the weaker 
5854, 6141 and 6496~\AA\ Ba~II lines are picked up as well.
Caution is necessary for these particular elements, but we believe that the 
magnitude of the
potential errors is sufficiently small that the fundamental conclusions of 
our work are not affected.

\section{Abundance Ratios \label{section_ratios}}

\subsection{C/H Ratios}

We have two indicators in the present work for the carbon abundances,
the strength of the bands of CH and of \cband.
The upper panel of Fig.~\ref{fig_cnratio} shows $\epsilon$(C) inferred from the
G band of CH
as a function of [Fe/H](HIRES) for the full
sample of \ncstars\ C-stars and the three EMP C-enhanced stars with
[C/Fe] $> 1.0$ dex from our work. Eleven additional
very metal-poor C-stars, mostly from the HK Survey,  
with recent analyses from the literature, are indicated
as small open circles in this figure as well as in
Fig.~\ref{fig_carbon} and in Fig.~\ref{fig_ba_c_feh}.
The details for the
additional stars are given in  Table~\ref{table_add}.  This produces
a total sample of
27 Fe-poor C-stars and three EMP C-enhanced dwarfs.

The dashed horizontal line in the upper panel of Fig.~\ref{fig_cnratio}
indicates a constant C/H ratio
of 20\% of the solar value independent
of [Fe/H]. This constant $\epsilon$(C), which we denote as $\epsilon_0$(C),
is a reasonable fit to
all the available data, given the uncertainties. The
inferred $\epsilon$(C)
reach a maximum value of $\sim$1/3 Solar, consistent with $\epsilon_0$(C).
Fig.~\ref{fig_cnratio} shows that EMP C-stars, even though they are of very low
[Fe/H], can, by whatever processes are relevant, achieve C-enrichment
up to near the Solar abundance, but not beyond it.  This is also true
of the two known 
ultra-metal-poor stars \citep{christlieb04,frebel05}.
\cite{marsteller} also have reached similar conclusions using the HK Survey sample.

The most metal-poor star shown in this figure is G77--61,
with [Fe/H] about $-4.0$ dex.  This star is a M dwarf 
in a binary system.  Since the star is so cool compared to the
C-stars studied here, it has a much more complex spectrum
with very strong molecular features. 
As part of a recent study by \cite{plez05} a search was made
for a detectable feature of O in the optical spectrum of this star.
However, given the very strong molecular bands in this M dwarf,
none could be found even in high precision Keck/HIRES spectra
covering the full optical spectral regime from 0.4 to 1.0$\mu$.
Thus in the abundance analysis for G77--61 carried out 
by \cite{plez05}, it was assumed that O was 
enhanced by +0.3 dex (i.e. [O/Fe] = +0.3 dex). The resulting
enhancement of C was found to be [C/Fe] +2.6 dex.
A recently obtained Keck high resolution near IR spectrum
yielded a detection of CO, and hence enabled determination
of the O abundance.  \cite{plez06} found an unexpectedly
high O-enhancement, [O/Fe] about +2.2 dex,
much higher than the previously assumed value.  With the original
value for $\epsilon$(C), this star would not be an extreme C-star, which 
its spectrum clearly demonstrates that it is.
The new higher O abundance thus in
turn led to a revised [C/Fe] value of +3.2 dex.
The values plotted in the figures for this
star (which is included in the additional sample from the literature) 
are these updated values.

In this context it is important to note that we also in general
lack a determination of the O abundance for
the C-stars in our sample (although clearly near IR
spectra of the CO bands would yield such), and have assumed
[O/Fe] to be the maximum of +0.5 dex or ([C/Fe] -- 0.8 dex)  
in calculating the molecular equilibria for all the
C-stars analyzed here. Only one star in our sample has a measured 
O abundance;
HE1410$-$0004 has [C/Fe] +2.0 dex, [O/Fe] + 1.2 dex, and
C/O = 5.  This O abundance
is in accord with the assumption we have chosen to make regarding [O/Fe] 
when no O abundance in available.  If the O abundance
in this star is in fact even lower,
which it might be given the marginal detection of the strongest line
of the 7770~\AA\ IR triplet, the molecular equilibrium for
CH and for \cband\ would not change significantly.
We expect the largest change in the deduced C abundance
(i.e. the largest shift in the molecular equilibrium of CH and \cband)
for C-stars as the O abundance is increased from that of C-normal stars
to occur when $\epsilon$(O) is only slightly less than
$\epsilon$(C). (Recall that $\epsilon$(O) must be less than 
$\epsilon$(C) since these are C-stars.)  For changes in $\epsilon$(O)
from the nominal value for HE1410$-$0004
given in Table~\ref{table_abundc}
not exceeding a factor of 4, the change in $\epsilon$(C) deduced from
the CH band in this star is modest, less than $\pm$0.15 dex.

\begin{figure}
\epsscale{0.65}
% Comment out the following line to embed the PS figure into the manuscript
% \plotone{/scr2/jlc/hamburg_survey/paper_cstars_2005/c2band_teff.ps}
\plotone{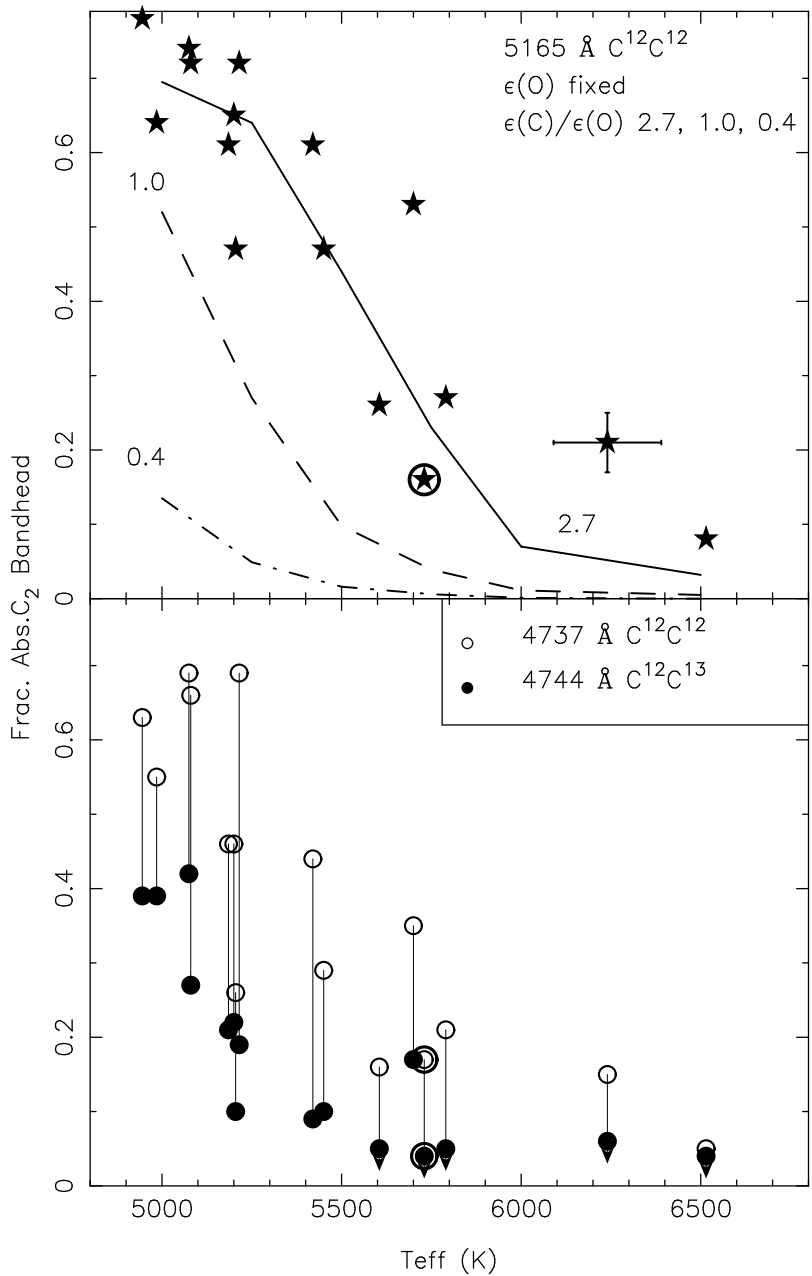}
\caption[]{The fractional absorption at several C$_2$ bandheads 
measured from our Keck/HIRES spectra
is shown for our sample of C-stars from the HES as a function
of \teff.  
The symbols are those of Fig.~\ref{fig_ion_eq}. 
Upper panel: the fractional absorption at 5165.0~\AA\ 
(the 0,0 bandhead of the Swan system) is shown.  The solid curve indicates
the predicted behavior for log[$\epsilon$(C,N,O)] = 7.56, 6.55, and 7.13 dex
(C/O 2.7),
the dashed curve is that for C/O=1 and the dot-dashed curve for
C/O 1/2.7, keeping $\epsilon$(O) fixed.
Lower panel: the 4737~\AA\ bandhead of $^{12}$C$^{12}$C is shown as open circles
while the 4744~\AA\ bandhead of $^{12}$C$^{13}$C is shown as filled circles.
Vertical lines connect the two values for each C-star.
All C-stars in our sample hotter than 5500~K only have upper limits for the latter.
\label{fig_c2band}}
\end{figure}

\begin{figure}
\epsscale{1.0}
% Comment out the following line to embed the PS figure into the manuscript
% \plotone{/scr2/jlc/hamburg_survey/paper_cstars_2005/c12c13_plot.ps}
\plotone{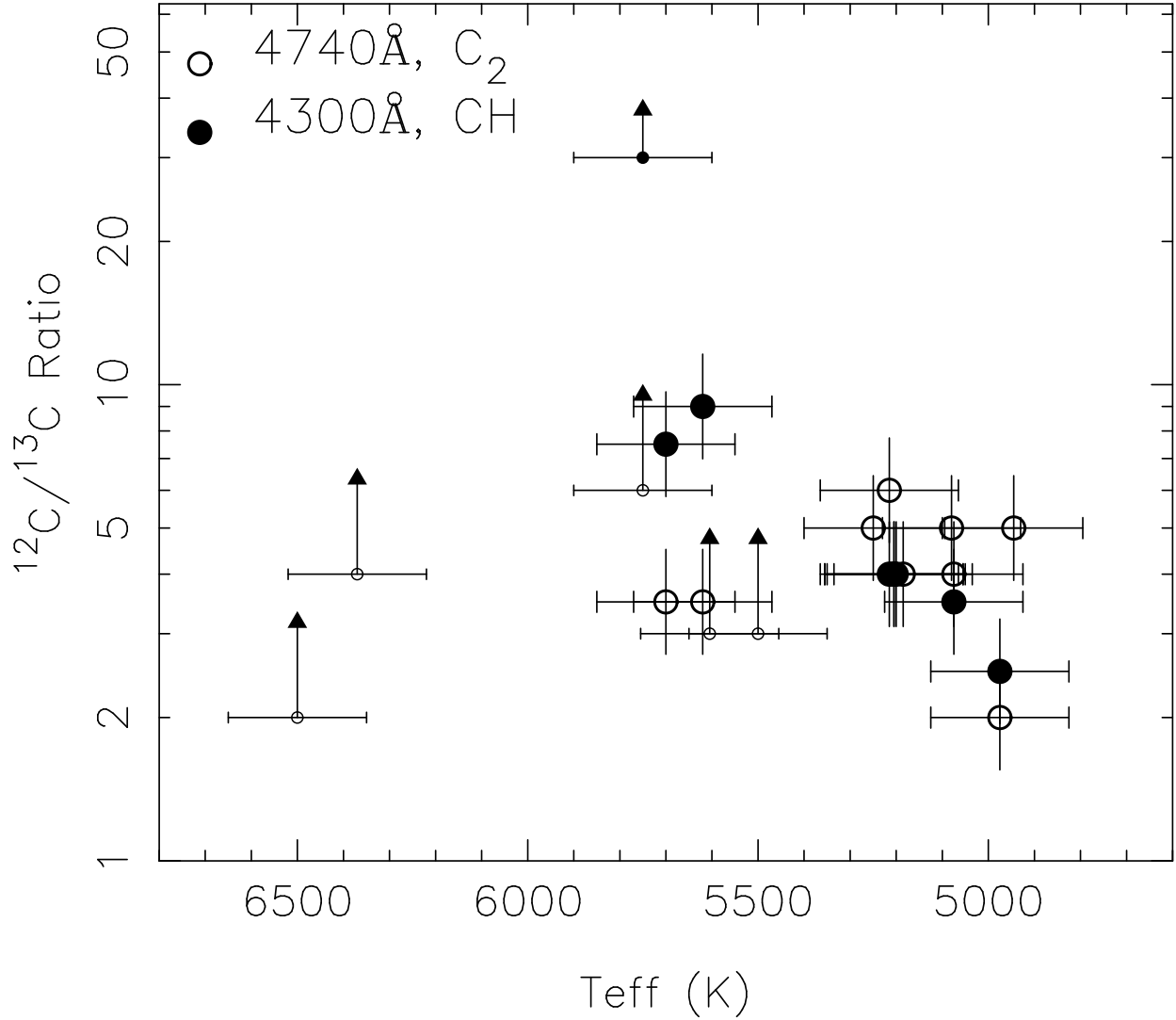}
\caption[]{\ciso\ ratios measured from the \cband\ (1,0) Swan band
and from the G band of CH
are shown as a function of \teff\ for the C-stars in our sample.  All 
C-stars with \teff\ $>$ 5700~K have only lower limits to the \ciso\ ratio
from the present spectra.
\label{fig_c12c13}}
\end{figure}

\begin{figure}
\epsscale{0.7}
% \figurenum{3}
% Comment out the following line to embed the PS figure into the manuscript
% \plotone{cfe_cnratio.ps}
\plotone{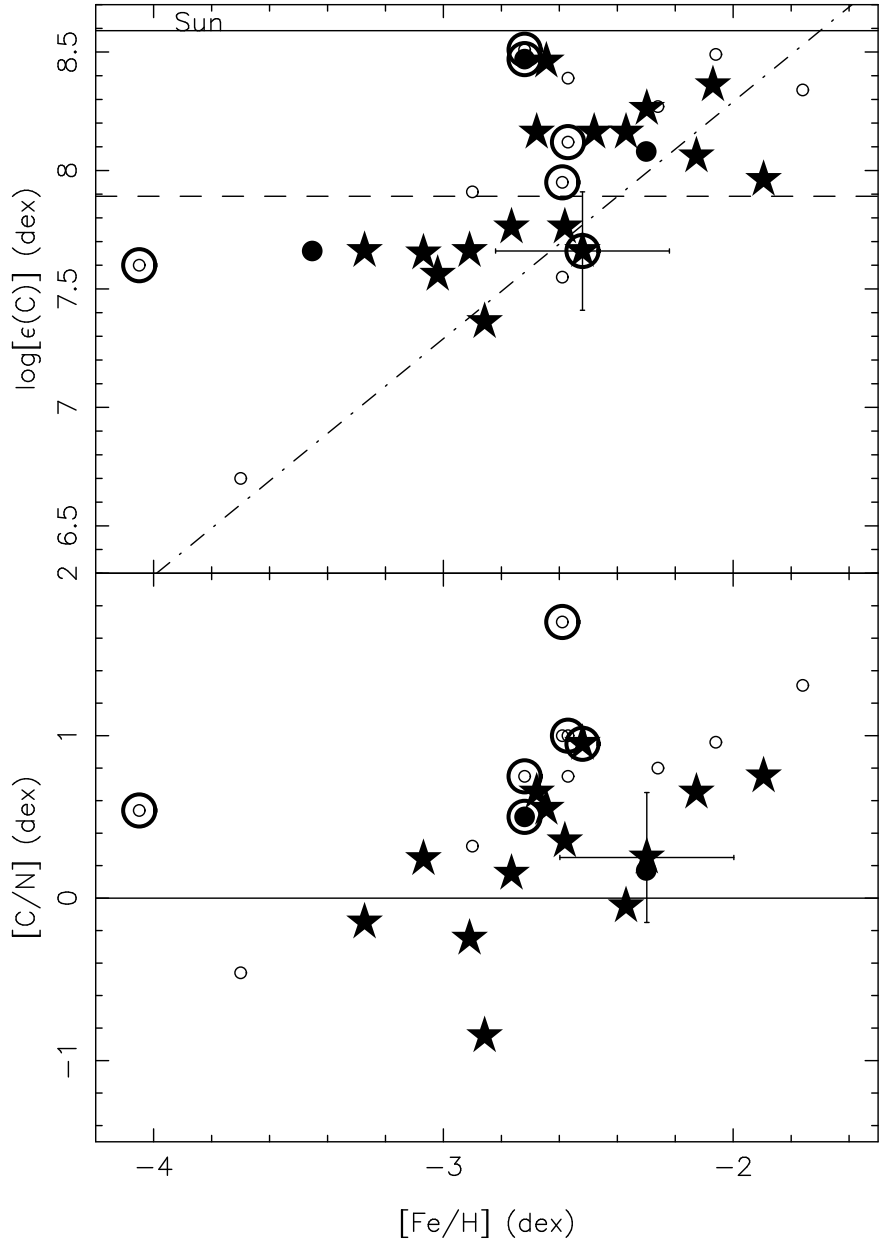}
\caption[]{Upper panel: log$\epsilon$(C) is shown as a function of [Fe/H] 
for C-stars (marked by large stars) 
and the C-enhanced stars (indicated by filled circles) 
with detailed abundance analyses from the HES by our group. 
The augmented sample of very metal-poor C-stars from the literature
(see Table~\ref{table_add} for details) is shown as small open circles.
Known spectroscopic binaries
are circled.  The dashed horizontal line indicates
a fixed $\epsilon$(C) of 20\% that of the Sun.  The sloping line indicates
the locus of [C/Fe] = +1.7 dex.  Lower panel: the
same for [C/N].  The horizontal line indicates the Solar ratio.
A typical error bar is indicated for a single star in each panel.
\label{fig_cnratio}}  
\end{figure}

\begin{figure}
\epsscale{1.0}
% Comment out the following line to embed the PS figure into the manuscript
% \plotone{/scr2/jlc/hamburg_survey/paper_cstars_2005/carbon_teff.ps}
\plotone{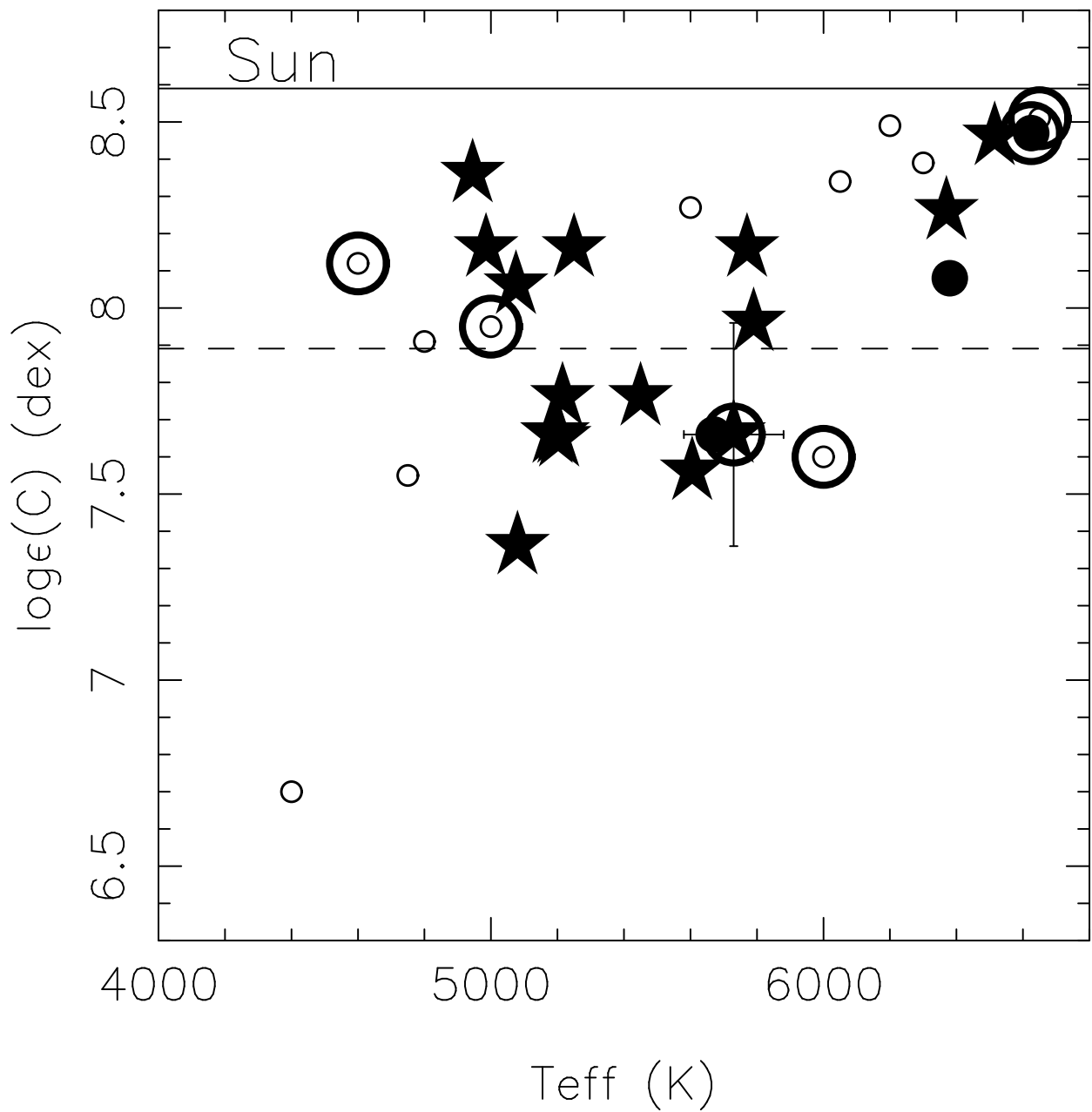}
\caption[]{log[$\epsilon$(C)] is shown as a function of
\teff\ for C-stars (large stars)
and C-enhanced stars (large filled circles) from our sample.  The 
augmented sample
of very metal-poor C-stars with recent detailed abundance analyses from 
the literature
(see Table~\ref{table_add} for details) is shown as small open circles.
Known spectroscopic binaries are circled.
The solid horizontal line represents the Solar ratio, while the horizontal
dashed line represents 20\% of Solar.  G77--61 is plotted as a
dwarf with \teff\ = 6000~K.
\label{fig_carbon}}
\end{figure}

The interpretation of the CH band strengths as a 
measure of the C abundance in the sample C-stars
is straightforward, ignoring the  issue
of the linkage to the assumed O abundance
discussed above.  With regard to \cband, 
we look again at the upper panel of  Fig.~\ref{fig_c2band} (a plot
of the absorption at several bandheads of \cband\ versus
\teff).   Although \cband\ band strengths
were not used to determine the
C abundance, spectral synthesis in the region of the 5160~\AA\ bandhead
with the fixed CNO abundances log[$\epsilon$(C,N,O)] 7.56,  6.55, and  7.13 dex
(a C/O ratio of 2.7) 
(for the value of $f_{00}$, the
band oscillator strength, adopted by Querci and collaborators) were used
to predict the depth of absorption at the 5160~\AA\ band head. 
The \teff, \grav\ pairs
were chosen to follow the isochrone for an age of 12 Gyr with
[Fe/H] $-2.5$ dex.  The result is shown as the solid curve
in the upper panel the figure, and clearly indicates that 
increasing absorption at the
\cband\ bandhead as \teff\ decreases is due to the
shift in the molecular equilibrium with \teff. 
Additional curves in this figure are shown for a C/O ratio of 1.0 and of 1/2.7,
keeping $\epsilon$(O) fixed, as would occur in a star to which C-rich
material is added.  The rapid decline in the strength of absorption at the
\cband\ bandhead is obvious and is due largely to the dependence of
$\epsilon$(C$_2$) on  $\epsilon$(CI)$^2$.  Our ability to match the
observed strength of the \cband\ bandhead in our sample of C-stars
shown in Fig.~\ref{fig_c2band} by
varying only \teff\
is consistent with the key result from analysis of the G band
of CH that
an approximately  constant $\epsilon$(C) is a satisfactory fit to the 
existing data on highly C-enhanced stars.

There are no stars in the upper right area of Fig.~\ref{fig_c2band}.  
This is,
in terms of \cband\ band detectability, an allowed area.
Thus
sufficiently strong bands of \cband, equivalent to sufficiently
large C-enhancements, do not exist in real stars with \teff\ $\sim$6200~K
with their higher continuum flux.
The required very large C-enhancements
in such hot stars must substantially exceed the 
constant $\epsilon_0$(C) deduced from the CH analysis.
The maximum \cband\ band strength, presumably that corresponding
to $\epsilon_0$(C), is very weak among the hotter
stars in our sample (the main sequence turnoff region stars), and
so stars with lower C-enhancements will simply have no detectable
\cband. One might  wonder why no stars appear in the lower left
corner of this plot, where weaker \cband\
features could easily be detected. This appears to be a consequence
of the fact that a C-star must have $\epsilon$(C) $> \epsilon$(O),
otherwise oxides will dominate the molecular equilibrium.  
At the solar composition, $\epsilon$(C)/$\epsilon$(O) is about 1/2.
Normal-C EMP unevolved and hence unmixed stars (i.e. low luminosity giants 
or dwarfs) have [O/Fe] about +0.7 dex, while they have [C/Fe] about +0.4 dex
\citep[see, e.g. for the giants][]{spite05}.  
A  C-enhancement of a factor of four for a normal-C unmixed 
EMP star will lead to  $\epsilon$(C) = $\epsilon$(O), and that required
to produce a C-star must be slightly higher.  The
\cband\ becomes stronger as the C-enhancement increases above the minimum
required to produce a C-star.  We
suggest that the duration of this phase of C-enhancement is short
compared to the age of the EMP C-star, and that this phase did
not in general occur recently as compared to the timescale for
mixing, making this region of Fig.~\ref{fig_c2band} unpopulated.

Among more highly evolved EMP and VMP C-stars, we would expect to see some 
evidence for depletion of C at the stellar surface as a result of
mixing and dredge up, which will depend on
the mass included in the mixing region.  We use \teff\ as a surrogate
for evolutionary stage, as the star cools as it moves up the RGB;
the M dwarf and EMP star 
G77--61 is plotted as though its \teff\ were 6000~K to place it
at the proper position corresponding to its evolutionary state in this figure.
Fig.~\ref{fig_carbon}
displays log[$\epsilon$(C)] as a function of \teff; the 11 additional
Fe-poor C-stars from the literature are included.  There is
a suggestion in this figure of decreasing $\epsilon$(C)
as \teff\ decreases, i.e. as the star moves up the giant branch,
reminiscent of mixing and dredge up phenomena
studied among EMP giants by \cite{spite05} and among globular
cluster giants by \cite{cohen05b}.  The slope of a linear
fit to the data in this figure is statistically different from 0.0
at more than the 3$\sigma$ level.
The existence of such a correlation, should further work demonstrate
conclusively that it is real, would again suggest that the C-enhancement could
not have occurred recently; sufficient time for C-depletion and mixing in the
giant EMP C-stars is required.

It is interesting to note that the highest value of \ciso\ we measured was
obtained using the G band of CH for the hottest and least luminous 
(and presumably least evolved) of the C-stars with a high 
signal-to-noise ratio HIRES spectrum.  
\cite{ryan06} compiled
\ciso\ ratios for Fe-poor C-rich stars from the
literature.  Their compilation also supports the suggestion
that there is a general
trend of declining \ciso\ with increasing luminosity.  This trend,
which needs further confirmation,
together with the generally low \ciso\ ratios,
is reproduced by the models of \cite{boothroyd99} as a consequence
of deep mixing and ``cool bottom processing'' after the first
and second dredge up in low mass red giants.  They establish that
the latter increases dramatically as [Fe/H] decreases.
Additional determinations of \ciso\ for EMP C-stars
from the \cband\ bandhead at 4740~\AA\
will be straightforward, and are now underway.

\subsection{Abundance Ratios for Other Elements}

Table~\ref{table_ratios} gives statistics for selected abundance
ratios for the sample of \ncstars\ C-stars from the HES analyzed here.  
Upper limits are ignored.   Only the sample of 16 C-stars analyzed here
is used to compute the statistical measures
given in Table~\ref{table_ratios}.  The mean abundance ratios
for various elements are compared with those obtained by
\cite{cohen04} for a large sample of EMP dwarfs, and in some cases
to those from the First Stars project at the VLT for EMP
giants \citep{cayrel03,spite05}.

The median [C/Fe] is +1.9 dex,
with a small dispersion (0.3 dex) about the mean.  The lower limit of 
$\epsilon$(C) is defined by the requirement that 
the star be a C-star to be included in the present sample,
but the upper bound is not constrained; it is
determined by the stellar characteristics
themselves.  
N is also highly enhanced, with a median [N/Fe] of +1.7 dex, only  slightly
below the median C-enhancement.
The  scatter is perhaps slightly
larger than that seen for $\epsilon$(C) in  Fig.~\ref{fig_carbon}.
The lower panel of Fig.~\ref{fig_cnratio} shows [C/N]
as a function of [Fe/H]. 
The mean is somewhat higher than the Solar value, but there
is no obvious trend of C/N with [Fe/H].  
Among the giants, there is a  suggestion that $\epsilon$(N) increases
and [C/N] decreases
as \teff\ decreases and luminosity along the giant branch increases,
but the scatter is large and this may not be statistically significant.

We include in our analysis two of the Mg triplet lines, which lie 
in a region free of molecular features.  Hence the Mg abundance
should be reliable\footnote{There is a minor
caveat regarding
the issue of internal consistency of the $gf$ values
between the various Mg lines discussed in \cite{cohen04}, but this
is a small effect, $\sim$0.2 dex at most.}.  The median abundance ratio [Mg/Fe]
of our C-star sample agrees well with that of the EMP dwarfs from \cite{cohen04},
but the
range of derived [Mg/Fe] is quite large (a factor of 10).  The highest
value, [Mg/Fe] = +1.04 dex (for HE0336+0113), is comparable to that of 
the small number of
other extremely Mg enhanced C-rich stars known, 
i.e. CS 29498--043 discussed by \cite{aoki02b} and
BS 14934--002 \citep{aoki05}.  The lowest value
of [Mg/Fe]
among the C-stars in our sample (+0.04 dex, for HE0212--0557) 
is comparable to the lowest seen
among VMP and EMP stars (see, e.g. the compilation in Fig.5 of
Aoki \etal\ 2005). [Mg/Fe] almost certainly shows a real range from star-to-star
among EMP stars.

The abundance of Ti should be well determined as there
are many strong Ti~II lines in the spectra of these C-stars, some of which
lie in regions completely free of molecular contamination.
Cr benefits from the
strong line at 5206~\AA, again a region unaffected by
molecular features.
It is thus gratifying that the [Ti/Fe] and [Cr/Fe] abundance ratios among the
C-stars from the HES show relatively small dispersion, with mean values in
good agreement with the results for EMP dwarfs from \cite{cohen04}.  
The remaining elements up to the Fe-peak suffer from a paucity of unblended 
lines with strengths sufficiently large for a reliable abundance analysis.

%The other abundance ratios for elements up to the Fe-peak
%in general suffer from species having one to a few 
%relatively weak lines in regions
%where molecular absorption may also be occurring.

We find that \ncbahigh\ of our C-stars show an enhancement of Ba
(see Fig.~\ref{fig_bafe_feh}) and other $s$-process
neutron capture heavy elements approximately equal to that of C.
The other four  show [Ba/C] $\leq -1.6$ dex, i.e. a strong
C enhancement, with normal heavy elements, as contrasted to
enhancement of both C and the $s$-process elements in the majority
of the C-stars.  In the full sample of 27 C-stars and three C-enhanced
dwarfs, 6 stars do not show
a strong Ba enhancement, while $\sim$85\% of the full sample
do show a strong Ba-enhancement.
Fig.~\ref{fig_ba_c_feh} shows the [Ba/C] ratio for our sample of
HES EMP stars.  There is a strong
suggestion that the stars with low [Ba/C] ratios are the most
Fe-metal-poor of the sample.  The bifurcation into $s$-normal and highly C-enhanced stars
is not an artifact of relying on the Fe-abundances, which 
are decoupled from the C-abundances.

We can examine whether the process that produces highly enhanced C in these C-stars also
leads to abnormalities in the abundances of other elements beyond those
established above, i.e.
CNO and the heavy elements beyond the Fe-peak.  
We define $\Delta(X)$ as 
the difference between the median [X/Fe] in our
C-star sample with HIRES abundance analyses and that found
for C-normal EMP dwarfs and giants.  From the values given
in Table~\ref{table_ratios}, for elements 
from Na to Fe we find only
two with $\mid\Delta(X)\mid ~ > ~ 0.25$ dex.  These are Al ($\Delta$(Al) = +0.36 dex)
and Mn ($\Delta$(Mn) = +0.38 dex).   There is only one reliable line 
for Al~I (at 3961~\AA)
and only two for Mn~I (two of the three lines of the 4030~\AA\ triplet,
ignoring a few very weak lines of Mn which are only rarely detected
in the HIRES spectra of these C-stars) and each of these is
located  in regions of strong CH absorption.  It is likely that there
is still some contamination of the atomic features by molecular ones
that we were not successful in removing.  With this caveat,
we thus conclude that the C-star phenomenon in EMP stars 
is confined to the elements CNO and to the elements heavier than the Fe-peak.
The abundance ratios [X/Fe] of elements from Na to Fe for which we can detect
suitable lines are normal.

\subsection{Evidence that $s$-process Neutron Capture Dominates Among
the EMP C-stars}

We discuss here the evidence that enhancement of the neutron capture
elements seen in EMP C-stars arises from the $s$-process, with 
no substantial/detectable contribution 
from the $r$-process.
When we look at the elements beyond the Fe-peak, we notice
that the median and the mean [Eu/Ba] (both about $-0.8$ dex)
closely correspond  to that characteristic of
the main component of the solar $s$-process given by
\cite{arlandini99}.  The detection of large
amounts of lead is another clue that the $s$-process is responsible.
The median value of [Pb/Ba] 
(+0.79 dex, with $\sigma$ about the mean of 0.34 dex) is 
close to that of other $s$-process dominated stars:
\cite{sivarani} has compiled all the data for Pb in such stars
available to date (their
Table 5 and Figure 11).

Additional abundance ratios give clues to the detailed behavior
of the $s$-process.  For example, we find
a smaller range in [Y/Fe] and in [Sr/Fe] than in [Ba/Fe],
which shows a range of a factor of 1000;
this is consistent with metal-poor $s$-processing in AGB stars.
\cite{busso99}, for example, predict
the $s$-process enhancement will be relatively larger 
for the second peak elements than for the
lighter $s$-process nuclei in stars with lower Fe-metallicity.  
A recent extensive theoretical discussion
of the nucleosynthesis of Sr, Y and Zr was given by
\cite{travaglio04}.

Ignoring the upper limits, $\sigma$[Y/Sr] and $\sigma$[La/Ba]
are small (0.32 and 0.26 dex respectively), confirming
previous work suggesting that within each of the 
peaks, the $s$-process element ratios 
for the Ba-rich EMP C-stars are approximately constant
for elements within that particular peak,
while the variation from star-to-star of the
ratio of the strength of the various peaks is much larger.
\cite{aoki05} also present relevant data for
a sample of 18 very metal-poor stars supporting this.

\subsection{The Ba-poor C-stars \label{section_bapoor} }

\begin{figure}
\epsscale{1.0}
% Comment out the following line to embed the PS figure into the manuscript
% \plotone{/scr2/jlc/hamburg_survey/hires_summary/programs/bafe_feh.ps}
\plotone{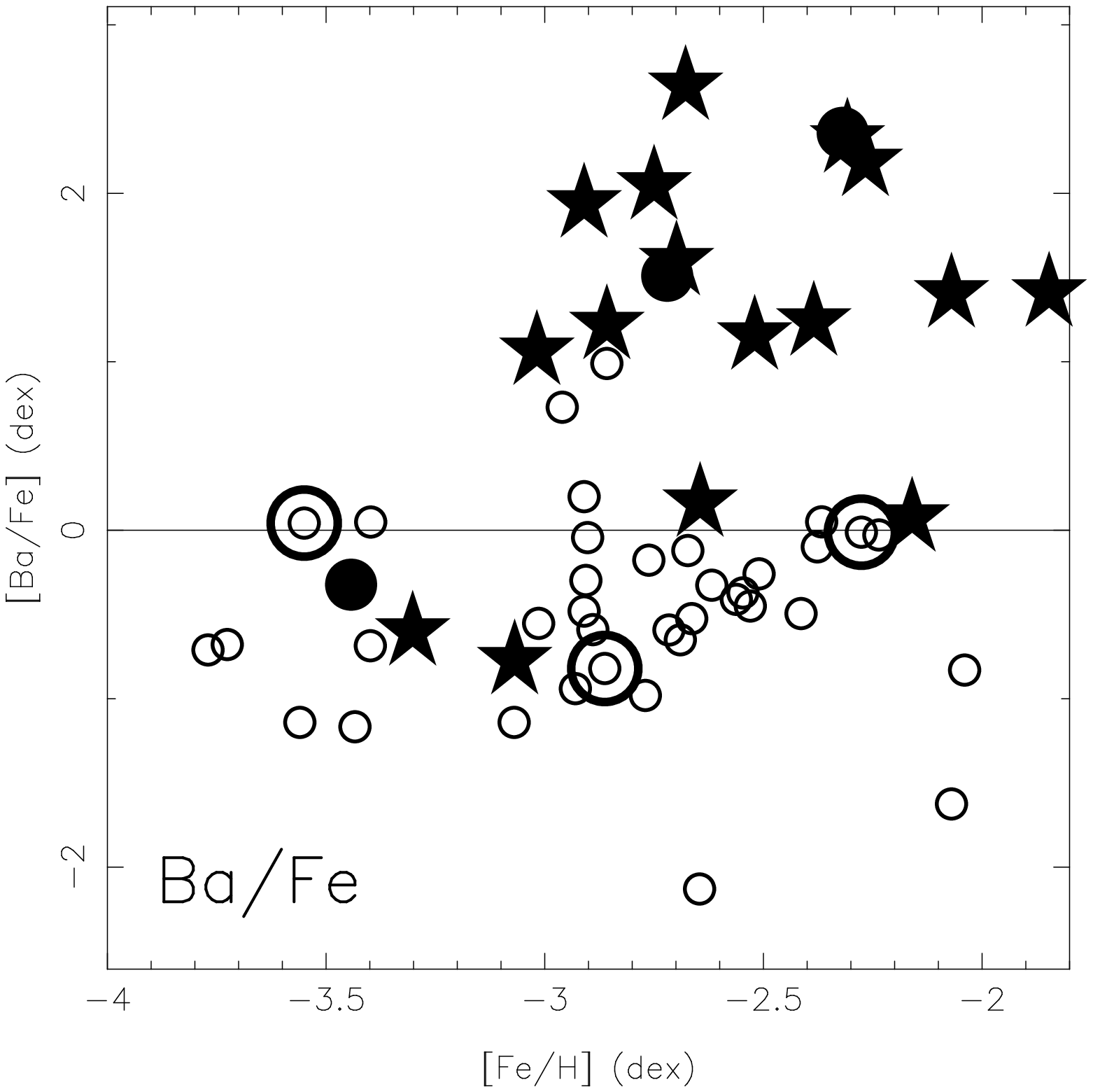}
\caption[]{The abundance ratio [Ba/Fe] as a function of
[Fe/H] for HES EMP C-stars (large stars) and 
C-enhanced stars (large filled circles) with detailed abundance
analyses.  All C-normal stars from the HES analyzed to date by us
are shown as small open circles.
The additional  C-stars from the literature are not shown.
Known 
spectroscopic binaries are circled. 
\label{fig_bafe_feh}}
\end{figure}

\begin{figure}
\epsscale{1.0}
% Comment out the following line to embed the PS figure into the manuscript
% \plotone{ba_c_feh.ps}
\plotone{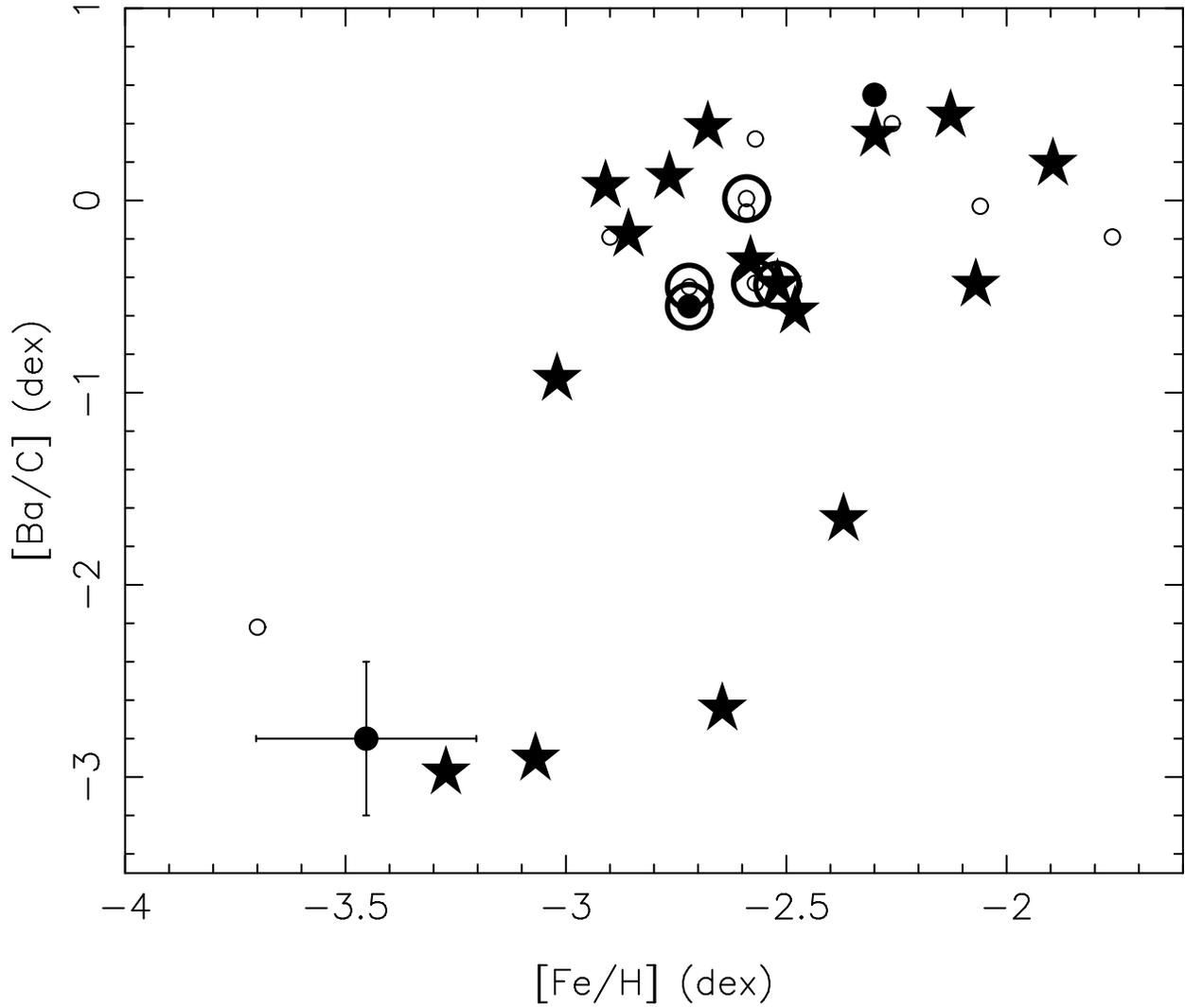}
\caption[]{The abundance ratio [Ba/C] as a function of
[Fe/H] for HES EMP C and C-enhanced stars with detailed abundance
analyses and for the augmented sample of very metal-poor C-stars from 
the literature. 
The symbols are those used in Fig.~\ref{fig_carbon}; known 
spectroscopic binaries are circled.  A typical error bar is indicated
for a single star.
\label{fig_ba_c_feh}}
\end{figure}

Fig.~\ref{fig_ba_c_feh} shows [Ba/C] as a function of Fe-metallicity for this
sample of C and C-enhanced stars.  
Just as was seen in Fig.~\ref{fig_bafe_feh},
% \ncbahigh\ 
12 of the C-stars from the HES that we have analyzed
show an enhancement of Ba (and of the other $s$-process
neutron capture heavy elements) approximately equal to that of C.
The other four show [Ba/C] $\leq -1.2$ dex, i.e. a strong
C enhancement, with more normal heavy elements.  Including 10
additional C-stars compiled from the literature, 25 of the 30
stars in the full sample of EMP/VMP C-rich stars (83\%) 
show highly enhanced Ba,
while 1/6  have [Ba/C] $\le -1.2$ dex.
It is clear from the evidence described above that
the $s$-process is responsible for the enhancement
of the heavy neutron-capture elements in these C-stars, 
when they are highly enhanced.
We note the Ba-poor C-stars that are cooler than \teff\ = 5700~K have 
the same low
\ciso\ ratios as do the Ba-rich C-stars.

We first consider whether the Ba in the Ba-poor stars is from the
$s$ or the $r$-process.  One might argue for the former, claiming
that Ba is in fact enhanced even in the Ba-poor stars.  
The influence of the very low Fe-metallicity on the
heavy neutron capture rates might give rise to a very low
$s$-process production, with the $r$-process making no
or an even lower contribution.  However, Fig.~\ref{fig_bafe_feh} shows
that [Ba/Fe] in the Ba-poor stars is consistent with that
observed among the C-normal stars from the HES that we have
analyzed to date.  We know that the Ba in C-normal EMP stars
must be largely produced in the $r$-process based on their
[Ba/Eu] and [La/Eu] ratios (e.g. McWilliam et al. 1995b, McWilliam 1997, 1998a,
Simmerer 2004). Thus we infer that the Ba in the Ba-poor EMP C-stars
has its origin in the $r$-process as well.

At first sight, the existence of two more or less distinct
classes of EMP C-stars suggests that two distinct processes are
required to produce the C-stars which are Ba-enhanced and those that are not
Ba-rich.  Nucleosynthesis within an intermediate mass AGB star can
reproduce the first set of characteristics.
If the mass of the EMP C-stars is assumed to be the turnoff mass of the halo
with an age of $\sim$12 Gyr,
near 0.8 M$_{\odot}$, they are not massive enough to produce s-process
elements at any time (e.g. see the review by Busso et al. 2004).  Also, their
\teff\ are too warm and the luminosities are too low for our C-stars to 
be AGB stars.  Thus intrinsic nucleosynthesis production and transport to 
the stellar surface of large amounts of C is not possible for such unevolved stars.  

We suppose instead that the EMP C-stars are the former secondaries of binary
systems across which mass transfer has occurred.  This is the mechanism originally suggested
for the CH stars by \cite{mcclure85}, which also have enhanced C and Ba 
and low Fe-metallicities (e.g. Wallerstein \& Greenstein 1964; Vanture 1992), 
although with $\epsilon$(Fe) still a factor of 50 to 100
times higher than the EMP C-stars discussed here, so the apparent enhancements
are not as large in the CH stars.
\cite{mcclure84} \citep[see also][]{mcclure90} established that essentially
all CH stars are members of binary systems.  The higher metallicity
Ba stars appear to be another example of the same phenomenon
\citep{mcclure90}; \cite{vitense} have established from UV HST spectra
the presence of white dwarf companions for several of these stars.

What about the 1/6 of the C-rich stars without heavy element enhancements?
We suggest that there is no need to resort to intrinsic production or any
other additional mechanism; in our view, essentially {\it{all}} of these stars 
could be produced by mass transfer and other phenomena in binary systems.
There are several possibilities for explaining these stars within
the context of our hypothesis that all EMP C-stars are or were binaries.
We can ascribe
the differing enhancement of the $s$-process elements
from C-star to C-star within our sample
to some dependence in the nucleosynthetic yields
involving, for example, the initial [Fe/H] or
mass of the original primary star.
At the lowest metallicities, 
\cite{busso99} (see especially their Fig.~12) predict that when 
$n(\rm{Fe~seed})$ becomes very small, there are so many neutrons
available for each seed nucleus that the $s$-process runs to completion,
with lead the main product, and very little Ba enhancement.
Lead is the third $s$-process peak, and $\epsilon$(Pb) is
considerably higher in the Sun than that of its neighbors in the
periodic table.  Any heavier elements produced, which are all unstable
except for Bi at atomic number 83,
decay to lead.
Although the prediction of \cite{busso99}
for the Fe-metallicity at which the peak Ba $s$-process
production occurs in AGB stars may be slightly too high,
their Fig.~12 shows a drop of more
than a factor of 100 for the predicted [Ba/Fe] enhancement
as [Fe/H] drops 1 dex lower than that at which maximum Ba production
occurs.  

We attempt to estimate the expected Pb abundance for a EMP C-star
assuming the $s$-process runs to completion
to see if it is detectable.
The highest [Ba/Fe] seen among the C-stars in our sample (see Table~\ref{table_ratios})
has $\epsilon$(Ba) approximately at the solar value for a C-star with [Fe/H] $-2.3$ dex.
We make the reasonable assumption
that $s$-process production is proportional to the number
of Fe seed nuclei, and assume that all the $s$- process elements
in the Sun, from Ba to Pb, end up as lead.    But all the intervening
elements have very low $s$-process abundances, see, e.g.,
the $s$-process solar abundances for the heavy
elements tabulated by \cite{burris00}.  Thus
for a [Fe/H]
$-3.5$ dex star, we predict $\epsilon$(Pb)
to be +1.5 dex. 
This Pb abundance, which is roughly 2.5 times the solar
Pb abundance, is a very high Pb abundance for such a low
Fe-metallicity star.  However, it
is, as discussed in \S\ref{section_lead},
extremely difficult to detect in a highly C-enhanced 
(recall that $\epsilon_0$(C) $\sim$1/5 solar) star with strong 
molecular bands given that the strongest Pb~I line at optical
wavelengths is weak  and located in a thicket of CH
features.
Thus verification of this idea through an abundance determination
extending to the third $s$-process peak will be very difficult in
practice.  We do, however, expect in this case that the 
Ba-poor EMP C-stars to be predominantly
those of the lowest Fe-metallicity, which does appear to be the case
in our sample 
(see Figures~\ref{fig_bafe_feh} and \ref{fig_ba_c_feh}),
in the somewhat smaller sample of \cite{ryan06},
as well as in that of W.~Aoki (private communication).

Another possible explanation for the absence of $s$-process enhancements
in some of our EMP C-stars is that the neutron flux is strongly reduced
in the AGB star, either due to low temperatures in the intershell region,
or because the $^{13}$C pocket fails to be injected into the intershell 
region of the AGB star, thus restricting the the $^{13}$C($\alpha$,n)$^{16}$O 
reaction. 
This $n$-producing reaction
competes with the reaction $^{13}$C($p,\gamma)^{14}$N.  At lower $T$,  
the latter may dominate, which would reduce the production
of neutrons available to create $s$-process elements.  The circumstances
which might lead to
lower $T$ are not clear, perhaps lower Fe-metallicity 
is in some way the dominant factor. 
In the absence of the neutron flux the $s$-process can not operate 
with vigor, thus producing the Ba-poor stars.

\begin{figure}
\epsscale{1.0}
% Comment out the following line to embed the PS figure into the manuscript
% \plotone{cmd_sample.ps}
\plotone{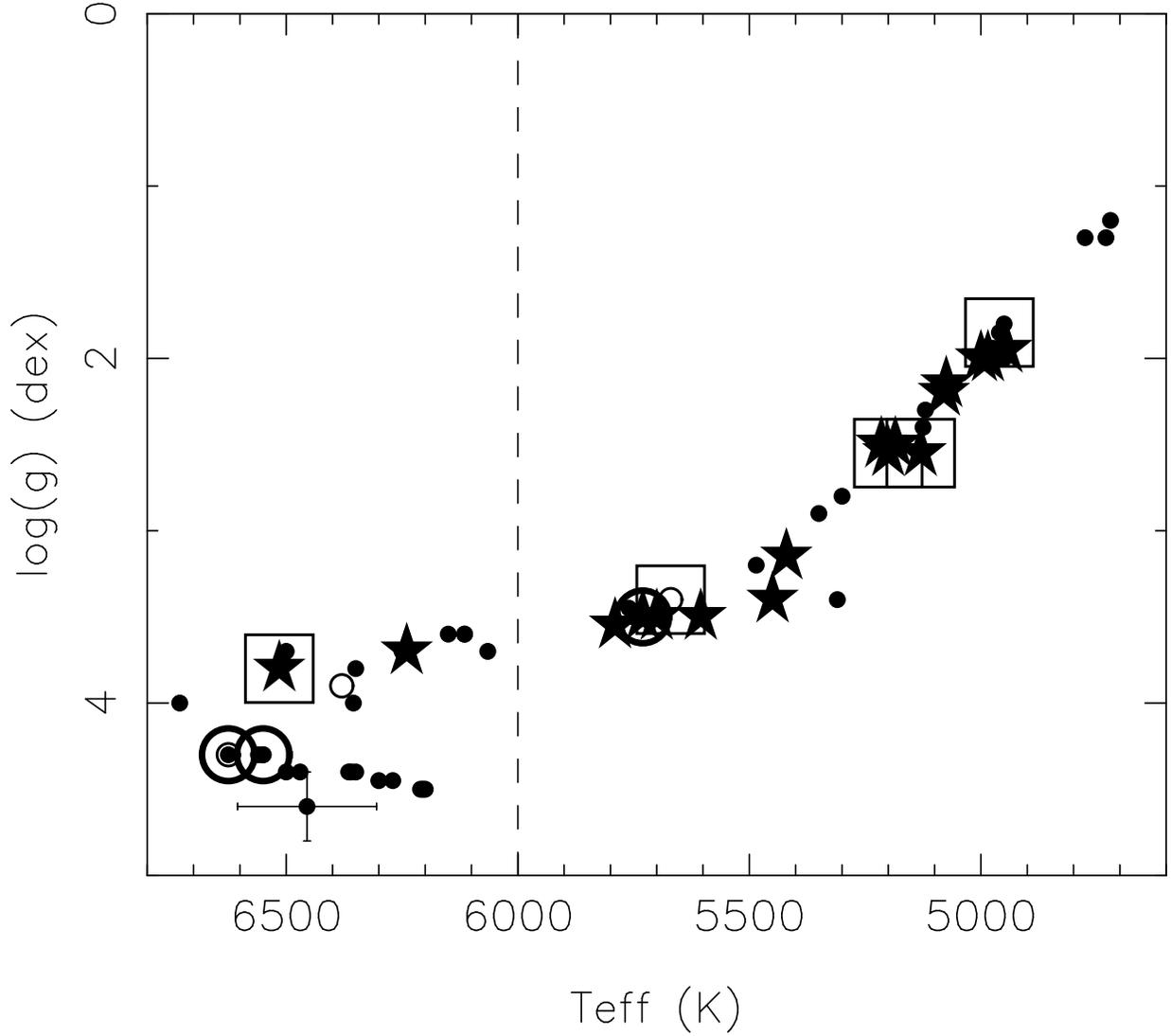}
\caption[]{The HR-diagram (\teff\ versus \grav) is shown
for our sample of 16 C-stars and 3 C-enhanced
stars denoted by filled stars for the former and by open circles
for the latter.  The small filled circles
indicate all the other EMP candidates from the HES that we have analyzed
to date.  The five Ba-poor, C-rich stars from our sample are enclosed
in squares.  The three known binaries are circled. 
The additional  C-stars from the literature are not shown.
A typical error bar is indicated
for a single star.
\label{fig_hr}}
\end{figure}

We view the trend for the Ba-poor C-stars to be among the most
Fe-poor as a fundamental clue to the
mechanism(s) involved in producing the Ba-poor C-stars.
Any differences in the luminosity distribution of the two groups of
C-stars might provide other useful clues for identifying the mechanisms 
involved.
Fig.~\ref{fig_hr} shows a \teff, \grav\ diagram for our sample
of C-rich stars.  Also shown there is the entire sample of
EMP candidates from the HES for which we have carried out 
detailed abundance
analyses to date.  Our sample is selected from the HES
and stars are chosen for HIRES observations and subsequent abundance
analyses solely on the basis of apparent low [Fe/H]\footnote{It must be
admitted that all the HIRES spectra of C-stars in hand as of Aug. 2005 have been
analyzed, but not all the spectra of C-normal stars in hand have
been analyzed yet.  This bias only affects the relative ratio
of C-rich to C-normal stars in  Fig.~\ref{fig_hr}, but not their
distribution along the locus.}.
Thus the distribution of stars, both C-rich
and C-normal, along  the locus they follow in the HR-diagram must represent
some folding of the volume surveyed by the HES given the luminosity
at each evolutionary stage, the IMF for EMP stars, and perhaps selection biases
within the HES.
The additional C-stars from the literature are not shown
in this figure as they come from various sources
and the selection criteria imposed for high resolution studies is not clear. 

Fig.~\ref{fig_hr} suggests that the C-stars of both types are concentrated
towards high luminosities, and are relatively rare among the turnoff
region stars.  We ascribe this to a selection effect, as 
the G band of CH 
becomes weaker and harder to detect
for such hot stars, even if the C-enhancement is very large.  The
\cband\ bands become even weaker under such circumstances. Such hot
stars can only be picked out as highly C-enhanced from a high resolution
study.  Fig.~\ref{fig_c2band} demonstrates the
weakness of the \cband\ band in the hot turnoff stars.
Low SNR moderate resolution spectra are inadequate to 
securely detect such
weak bands. This is
the case for the Ba-poor but C-rich star HE0007--1832 from our sample 
(this and the other Ba-poor C-rich stars are marked in the figure,
as are the known binaries)
which is a dwarf C-star whose
analysis was published in \cite{cohen04}.  
The somewhat hotter main sequence turnoff at a fixed age
for lower metallicity stars
(\teff\ at the turnoff becomes hotter by 150~K
when the Fe-metallicity decreases from $-2.2$ to $-3.2$ dex) makes
the CH and \cband\  bands in the lowest metallicity stars
near the main sequence turnoff even weaker
and harder to detect. 

\cite{ryan06}, in a very recent paper discussing the origin of the two
classes of C-enhanced metal-poor stars described above, postulate
two distinct mechanisms, with mass
transfer in an AGB phase of a binary system giving rise to the
Ba-rich and C-rich stars, while the 
Ba-poor, C-rich
stars are assigned a completely different origin.  
However, the discussion
given above indicates that there are several plausible scenarios 
for producing the Ba-poor
EMP C-stars within the framework of the binary hypothesis adopted here.   
We do not
find any reason at present  to exclude them from also being formed 
via phenomena
involving binary systems. 

The path to resolve the origin of the Ba-poor EMP C-stars, which
is in our view the only remaining area of considerable uncertainty
in our scenario, is difficult.
It requires assembling a larger sample of such stars, searching
with exquisite high resolution spectra for the presence of Pb, and
extensive radial velocity monitoring of these stars.

\subsection{Comparison with Disk C-Stars \label{section_disk}}

A comparison of the properties of the EMP C-stars with 
those having Fe-metallicity
near solar is of interest.
\cite{wallerstein98} present a review of the luminosities
and abundances of the latter.
Intrinsic C-stars 
stars which produce C internally, then dredge it up to the stellar surface,
are AGB stars with luminosities much higher than those of the EMP
C-stars in our sample.
\cite{lambert86} have analyzed such luminous cool disk C-stars;
their \teff\ is considerably lower than the stars studied here.
Their sample
has [Fe/H] $\sim -0.3$ dex, and shows only modest C-enhancements
(less than a factor of 2,
far smaller than the factor of $\sim$100 seen in our sample), with
no enhancement of N, and with \ciso\ typically large, 30 to 70.
The \ciso\ in these intrinsic C-stars 
suggests the addition of pure $^{12}$C from He burning, with
quite different abundance ratios among the CNO elements and 
also quite different \ciso\ ratios
than those seen among much more Fe-poor
C-stars studied here.  The difference between the C/N ratios
may arise if the former primary 
of the binary EMP C-stars in our sample had, in the mean, a different
stellar mass when it was on the AGB than is typical of disk
solar Fe-metallicity AGB stars, so as to produce different 
abundance ratios. Higher mass AGB stars produce higher C/N ratios. 
The predictions from the models of \cite{boothroyd99} are also relevant here, in that
a dependence of the nuclear reaction rates and hence
the internal production ratios
on [Fe/H] might also contribute to these differences.

It is now possible to investigate the abundances of C and N for luminous
AGB C-stars in the LMC and the SMC.  Preliminary
results by \cite{marigo}, \cite{matsuura} and \cite{vanloon} suggest
that the differences in abundance ratios between these more Fe-poor
luminous AGB stars and
Galactic disk intrinsic C-stars are small.  There is, however, a well known
decrease in mean luminosity and increase in the C-star to late M giant ratio
as [Fe/H] decreases from the Galaxy to the LMC and then to the SMC,
first discussed by \cite{blanco80}.  This presumably arises as a smaller
amount of C (of intrinsic origin; these are luminous AGB stars)
needs to be added to a very metal-poor star with a fixed
[O/Fe] ratio to reach $\epsilon$(C) = $\epsilon$(O) and so 
produce a C-star as [Fe/H] decreases.

The early R-stars (a type of C-star)
are much closer in some of their properties to the Ba-poor
EMP C-stars found in the HES that are studied here.  
\cite{dominy84} and \cite{dominy85} studied their chemical compositions
and evolutionary state. 
(See also the review of Wallerstein \& Knapp 1998.)  The R-stars 
are of lower luminosity than the intrinsic AGB C-stars, with
$M_{bol} \sim -0.3$~mag, $L/L$\subsun $\sim 100$, and,
with \teff\ $\sim 4600$~K, are warmer
than AGB C-stars.  Their space
density is too high for them to be stars in the He shell burning
phase of evolution \citep{scalo79}.
They have [Fe/H] $\sim$ solar,
with moderate C enhancements ($\sim$ +0.7 dex), 
and somewhat smaller N enhancements, 
but have $\epsilon$(O)
at the solar value.   They,
like the EMP C-stars, have low \ciso\ ratios.  The R-stars do not in
general show enhancements of the $s$-process elements.
\cite{mcclure97} has demonstrated, via extensive radial velocity
monitoring, that they do not appear to be binaries; he
suggested that they are coalesced binaries.

Among the various families of high Fe-metallicity C-stars, there
appears to be a correlation that the stars with highest \ciso\
are those which have strong $s$-process enhancements, while those
with the lowest \ciso\ have little or no enhancement of the elements
past the Fe-peak.  This correlation may be due to the variation
with $T$
in the rate of the reaction $^{13}$C($\alpha,n)^{16}$O, which provides
the neutrons required for the $s$-process to occur, as compared to that
of   
the reaction
$^{13}$C($p,\gamma)^{14}$N, which suppresses the production
of neutrons from $^{13}$C burning, or perhaps to some property
of the $^{13}$C pocket.

\section{Implications of the Mass Transfer Scenario for EMP C-stars
\label{section_implications} }

We explore here the consequences of our assumption that mass transfer
in binary systems produces all C-stars at all [Fe/H] whose 
luminosities are so low that they cannot be intrinsic C-stars.
The stars being discussed here are very metal
poor, so that by adding a small amount of processed material through
binary mass transfer, a large change in surface abundances
of the secondary star can be
produced, which will lead to
much more obvious changes in the star's spectral characteristics
than would occur at solar metallicity.
Furthermore the efficiency of the complex process of
binary mass transfer depends on the mass of the primary star, which affects
the mass loss rate, being
higher for higher AGB luminosities, i.e. higher mass of primary, within certain limits.
d$M$/dt may also depend on the metallicity if the mass loss is driven
by radiation pressure on dust grains.  For a given d$M$/dt of the AGB star,
the accretion rate onto the secondary
is a function of the binary separation  and other orbital properties.  The net
result may be a highly variable efficiency for fixed initial
[Fe/H] and the initial masses of the two components of the binary system.

A key result presented above is
the approximately constant C/H ratio, $\epsilon$(C) = $\epsilon_0$(C),
in the photospheres of the C-stars in our sample, which we derive
from our analysis of their CH and \cband\ bands.  This is presumably
a consequence of the primary nature of C production in AGB stars. We assume this
constant level extends to higher Fe-metallicity, although a slight
upward trend as [Fe/H] increases cannot be ruled out at this point
(see Fig.~\ref{fig_cnratio}).
We consider adding this constant $\epsilon_0$(C) to stars
of both higher and lower Fe-metallicity than those studied here.  
As [Fe/H] rises, the impact of adding additional C (accompanied
by additional H as well) is diluted.
If we assume that C-normal EMP stars have [C/Fe] +0.3 dex and 
[O/Fe] +0.7 dex and that the stellar photosphere of the star we currently
observe consists of equal amounts of its initial material and of
material accreted from its AGB companion, then at 
[Fe/H] $\sim -1.4$ dex, the star, with its additional C, will just
achieve $\epsilon$(C) = $\epsilon$(O) with the additional C-rich 
material.  This falls to $-2.0$ dex if the 
final photosphere contains 20\% accreted material.
More Fe-rich C-normal stars
cannot become C-stars through the mass transfer process with
our assumptions unless the accreted material comprises more than
50\% of the stellar photosphere.

In this scenario we thus expect
for higher Fe-metallicities
to see stars which are C-rich, but without \cband\ bands.  These
presumably correspond to the CH stars.  They occur in the right
Fe-metallicity range, and
essentially all of them were
shown by \cite{mcclure84} \citep[see also][]{mcclure90}
to be binaries. The frequency of C-stars
in the HES as a function of [Fe/H] to be given in \cite{cohen06cfreq}
provides further support for this hypothesis.
At still higher Fe-metallicities, the C-enhancement
becomes too small to be noticeable.  However, $s$-process production
is to first order a secondary process proportional to the number of 
Fe seed nuclei (i.e. to [Fe/H]).
Thus $s$-enhancement (i.e. the high 
levels of [$s$/Fe]) will still be
present at high Fe-metallicity, although the details of the nucleosynthesis 
may shift the relative production of the $s$-process nuclei 
towards the first peak at Sr (see, e.g. Busso \etal\ 1999).
Such stars
presumably correspond to the Ba stars, which are of higher Fe-metallicity
than the CH stars. According to \cite{mcclure90} \citep[see also][]{luck91}, the Ba stars are another
example of the same phenomenon of mass-transfer in binary systems.

The situation at lower Fe-metallicities was explored in \S\ref{section_bapoor}.
We expect, as described earlier, the $s$-process to run through to lead, which will
be extremely difficult to detect, with very low production of the more easily detected 
$s$-process elements such as Sr, Ba, La, etc.

The low \ciso\ ratios seen in these EMP C-stars, both Ba-enhanced
and Ba-poor, provides
another important clue.  They, combined with the high C/N ratios,
suggest that a two phase process is required.  First, mass transfer
across the binary system from a low Fe-metallicity AGB star 
with intrinsic production
of C (and hence a high \ciso\ ratio) occurs.  This is then followed
by a phase of mixing combined with ``cold bottom burning''
as described by \cite{boothroyd99} 
to produce the observed C/N and \ciso\ ratios.
\citep*[See][for a description of the consequences of 
this mixing process in more metal-rich C-normal field stars.]{carretta00}
Since the degree of C-depletion appears to depend on the
luminosity of the C-star we observe today, that part of the 
processing cannot have occurred in the donor star of the binary.

\subsection{Binarity}

We have suggested that all EMP C-stars (i.e. those with
$-4 \lesssim {\rm{[Fe/H]}} \lesssim -2$ dex) are the original secondary
stars of binary systems in which mass transfer occurred.  We have
further suggested that this mass transfer from an AGB primary can
produce the abundance anomalies we see among the EMP C-stars, specifically
the high enhancement of $s$-process elements among $\sim$85\% of these
C-stars.  Those VMP/EMP C-stars with low or no $s$-process enhancement
are cases where some factor, most likely the low Fe-metallicity of the primary,
while still producing, mixing
to its surface, and transferring to the secondary star ample
amounts of carbon, did not achieve such for the easily detectable
heavy neutron-capture element Ba.

We consider here whether the statistics
of binary detection among very metal-poor C-stars can support our hypothesis that
all of these C-stars were once binaries.  We expect 
most/all of them to still be binaries with 
(invisible) white dwarf companions.  The HES C-stars
of our sample are
themselves not suitable for this purpose. They were only 
recently discovered to be interesting stars, and
most have only been observed for a single epoch.  They are in general
faint for high dispersion spectroscopic analysis.  There were no
radial velocity monitoring programs for such stars until very recently.
Even so, we have already found three confirmed binaries in our samples
of candidate EMP stars from the HES.

So we look instead at the sample of additional C-stars from the literature.
These stars are in general brighter than the HES C-stars in our sample,
and they have been known as interesting objects for timescales
of several years to a decade, giving more opportunity for radial
velocity monitoring.  Table~\ref{table_add} indicates which of these
are known binaries and gives their periods and $v_r$ amplitudes. 
Four of these 11 C-stars are confirmed binaries,
consistent with the very preliminary results of
the $v_r$ monitoring program of
\cite{tsangarides} for $s$-process enhanced C-stars.

Although the sample is small,
considering the lack of suitable long-term radial velocity  monitoring programs,
the length of the typical period, the small velocity amplitudes,
the faintness of the stars, and the relatively short time they have
been known to be interesting, we find our detection rate for binaries
among very metal-poor and EMP C-stars to be consistent with
all such stars being binaries; Monte Carlo
simulations by \cite{lucatello} support this. 
There is as yet insufficient $v_r$ monitoring data for the small fraction of
C-enhanced stars without
$s$-process enhancement to assess their binarity.

\section{Summary}

We have studied a sample of \ncstars\ C-stars from the EMP candidate
lists of the HES using high dispersion spectra from HIRES at 
Keck and new optical photometry. 
We have carried out a detailed abundance analysis
using a \teff\ scale based on V--I, V--J and V--K colors, 
while avoiding the effects of the molecular
bands as much as possible.  Earlier \teff\ scale problems
affecting the Fe-metallicity deduced for EMP stars as hot as 6000~K
by the HES (and, until recently, the HK Survey) were solved
by changing from B$-$V to H$\delta$ as a \teff\ indicator.  
Our results provide a broad database to establish the Fe-metallicity for 
EMP C-stars.
We find that the Fe-metallicities 
for the cooler C-stars (\teff\ $\sim$5100~K) are still being underestimated
by a factor of $\sim$10 by the  current standard
HES (and until very recently HK) survey tools.
This is due to strong molecular absorption
primarily in the red continuum bandpass of the HP2 index which
measures the strength of H$\delta$ and acts as an indicator of \teff.
The results presented here provided crucial supporting data 
used by \cite{cohen06apjl} to derive
the  frequency of C-stars among EMP stars.

Carbon abundances in these very metal-poor stars appear to be constant,
independent of Fe-metallicity, at about 1/5 the solar value.  The
C-abundances
show marginal evidence of decreasing with decreasing \teff\ or increasing
luminosity, presumably due to mixing and dredge-up of C-depleted material.
Such C-depletion
is seen among ``normal'' halo field giants over a wide range of metallicity
for sufficiently evolved stars with
luminosities brighter than that of the RGB bump, which
is high on the red giant branch.  N is also highly enhanced in the EMP C-stars.
Among the elements studied here,
abundance anomalies in these stars appear to be confined to
CNO and to those heavier than the Fe-peak.

C-enhancement in this sample is associated with strong enhancement of
$s$-process heavy nuclei for \ncbahigh\ of the \ncstars\ stars, with [C/Ba]
about $-0.1$ dex with small scatter.  The remaining
four C-stars from the HES show no evidence for enhancement
of the heavy elements, with Ba providing the strongest constraint,
[Ba/Fe] $\leq +0.20$ dex 
for each of the four stars. 
When 11 additional C-stars, mostly from the HK Survey, with recently
published detailed abundance analyses are added, the same separation
is seen, with $\sim$85\% of the stars having [C/Ba] almost Solar.

Very high enhancements of lead are detected in some of the C-stars with 
highly enhanced Ba.
The ratio Ba/Eu, the high Pb abundances, and the high ratios of
diagnostic elements in the second to the first $s$-process peak
for C-stars 
in our sample demonstrate that the $s$-process 
is responsible for
the enhancement of the heavy elements for most of the C-stars in our
sample.  The mostly low \ciso\ ratios inferred from both the
G band of CH and the 4740~\AA\ band of \cband, where the isotope ratio is
particularly easy to measure, as well as
the high N-enhancements suggest that
the bulk of the stellar envelope of these stars has been processed
through the CN cycle of proton burning.
Our data for the Ba-rich C-stars supports the suggestion that
the abundance ratios for elements within a given $s$-process peak
are to first order  constant, while the ratio of the strength of the various
peaks shows larger star-to-star variations.

The similarities and differences
of the properties of the EMP C-stars to those of
various types of near solar [Fe/H] disk C-stars are discussed.
In particular, the early R stars show low \ciso\ ratios 
and no excess of the heavy elements, reminiscent of the
Ba-poor EMP C-stars found (at a low rate) in our sample.

The abundance ratios we derive are used
to discuss the origin of the C-rich stars among EMP stars. 
We suggest that both the $s$-process rich and Ba-normal
C-stars result from phenomena associated with binary
stars.  The  Ba-rich EMP C-stars presumably
formed as secondaries in a mass transfer binary system with an AGB
primary.   This was followed by proton burning at moderate
$T$ to reduce \ciso\ and increase the C/N ratio.
The implications of this hypothesis for stars of both higher and
lower Fe-metallicity than those in the present sample are
discussed.  Several possible origins for the 
small minority of Ba-poor EMP C-stars
are suggested.
In the most metal-poor stars, \cite{busso99} predict that
the $s$-process runs to completion through the
Ba-peak to the heaviest stable element, lead, leaving little or no
apparent Ba-excess.  Heavier elements (all unstable except Bi) mostly
decay to lead as well.  The predicted $\epsilon$(Pb) in a [Fe/H] $-3.5$ dex
star, while very high for a star with such a low Fe-metallicity,
will be very difficult to detect.
Another possibility for explaining the Ba-poor EMP C-stars
is a possible lack of neutrons
due to $^{13}$C burning via $^{13}$C($p,\gamma)^{14}$N instead of via
$^{13}$C($\alpha,n)^{16}$O.  The former dominates at lower $T$, 
while the latter provides
the neutrons required for the $s$-process to occur.  If either
of these suggestions is correct, the Ba-poor C-stars should 
have lower Fe-metallicities
in the mean than the Ba-rich C-stars, which does appear to be the
case in our sample.
The frequency of known binaries among the samples appears consistent
with our hypothesis for the origin of EMP C-stars 
given the lack of long term radial velocity
monitoring programs, the long periods, the low velocity amplitudes,
and other characteristics of the stars.

We thus see no reason at present to exclude the scenario adopted here, that
{\it{all}} the EMP C-stars are formed via phenomena involving
binary systems.   For old stars of low Fe-metallicity, several mechanisms
described above may
lead to C-stars with little or no $s$-process enhancement, such as is
occasionally
seen in our sample.  
For old stars  in binary mass transfer systems of higher [Fe/H] than
those considered here, a progression with 
increasing [Fe/H]  from C-stars to CH stars and finally to
Ba stars is predicted for a constant
donor $\epsilon_0$(C), which successfully reproduces several key
observed characteristics of the behavior of C-rich stars in the Galaxy.

\acknowledgements

The entire Keck/HIRES user community owes a huge debt
to Jerry Nelson, Gerry Smith, Steve Vogt, and many other people who have
worked to make the Keck Telescope and HIRES a reality and to
operate and maintain the Keck Observatory.  
We are grateful to the W. M. Keck Foundation for the vision to 
fund the construction of the W. M. Keck Observatory.
We are grateful to W.~Aoki for providing his HDS/Subaru spectra of selected C-stars
to verify our $^{13}$CH line list.  We thank G. Wasserburg for helpful
discussions and moral support.
This publication makes use of data products from the Two Micron All Sky
Survey, which is a joint project of the University of Massachusetts and
the Infrared Processing and Analysis Center/California Institute
of Technology, funded by the National Aeronautics and Space Administration
and the National Science Foundation. 
JGC is grateful for partial support from  NSF grant
AST-0205951.  She is grateful for funds from
the Ernest Fullam Award of the Dudley Observatory
for help in initiating this work.
N.C. and F.J.Z. acknowledge support from Deutsche Forschungsgemeinschaft
through grant Re~353/44. N.C. is also supported by a Henri
Chretien International Research Grant administered
by the American Astronomical Society.

{}

%
% Table1
% 
\clearpage
\begin{deluxetable}{lclr rcr}
\tabletypesize{\footnotesize}
\tablenum{1}
\tablewidth{0pt}
\tablecaption{The Sample of C-Stars Selected as EMP Candidates from the 
HES Mostly from the Palomar Sample
\label{table_sample}}
\tablehead{
\colhead{ID} & \colhead{Coords.} & 
\colhead{V\tablenotemark{a}} &  \colhead{I\tablenotemark{a}} &
\colhead{Julian Date Obs.} & \colhead{$v_r$\tablenotemark{b}} &
\colhead{[Fe/H](HES)} \\
\colhead{} & \colhead{(J2000)} &
\colhead{(mag)} &   \colhead{(mag)} &
\colhead{(-- 2450000)} & 
\colhead{(\kms)} & \colhead{(dex)}}
\startdata 
Obs. with HIRES \\
 HE0007$-$1832 &  00 09 52.8 $-$18 16 12 & 15.462 & 14.831 & 
        \tablenotemark{f} \\
 HE0012$-$1441 & 00 15 27.1 $-$14 24 37 & 16.358 & 15.704 &
       2547.8627 & +11\tablenotemark{g} & $-2.61$ \\       
 HE0058$-$0244\tablenotemark{c} & 01 00 53.0 $-$02 28 20 & 13.727 & 12.933 &
       2179.0096 & $-$68.4 &  $-2.81$ \\
 HE0143$-$0441  & 01 45 37.8 $-$04 26 43 & 16.382 & 15.724 &
       2547.0190 & +121.8 & $-2.94$  \\
 HE0212$-$0557 & 02 15 02.5 $-$05 43 23 &   14.70~ & \nodata &
       2544.9791 & $-230.6$ & $-3.45$  \\
 HE0336+0113  & 03 38 52.8 +01 23 08 &  14.955 & 14.110 &
       2179.0586 & +66.6  &  $-2.51$  \\
 HE1031$-$0020 & 10 34 24.1 $-00$ 36 09 & 14.296 & 13.338 & 
       3152.7297 & +69.7 &  $-3.70$  \\
\enddata
\tablenotetext{a}{Our photometry from ANDICAM images.}
\tablenotetext{b}{Heliocentric $v_r$.}
\tablenotetext{c}{Rediscovery of CS22183$-$015.}
\tablenotetext{d}{Rediscovery of CS22957$-$027.}
\tablenotetext{e}{$v_r$ only from the Mg triplet lines.}
\tablenotetext{f}{See \cite{cohen04}.}
\tablenotetext{g}{Double lined spectroscopic binary, $v_r$ variable}
\tablenotetext{h}{HIRES spectrum obtained too late for analysis here.  C-star with
very strong Ba~II lines.}
\end{deluxetable}

%
% Table 2
%
\clearpage
\begin{deluxetable}{lccc rrr}
\tablenum{2}
\tablewidth{0pt}
\tablecaption{Stellar Parameters and Observations\label{table_teff}}
\tablehead{
\colhead{ID} &
\colhead{\teff } &
\colhead{log($g$) } &
\colhead{$v_t$} & \colhead{Exp. Time} & \colhead{S/N\tablenotemark{a}} &
\colhead{Source/Follow Up} \\
\colhead{} &
\colhead{(K)} &
\colhead{(dex)} &
\colhead{(km/s)} & \colhead{(sec: HIRES)} 
}
\startdata
 HE0012$-$1441 & 5730\tablenotemark{d} & 3.5 & 1.6 &  10,800 & 75  & Magellan \\ 
 HE0058$-$0244  & 5620 & 3.4 & 1.6 &  4800 & 100 & Magellan  \\
 HE0143$-$0441  & 6240 & 3.7 & 1.6  & 9600 & 80 & P200 \\
 HE0212$-$0557  & 5075 & 2.15 & 1.8 & 6000 &  100 & P200 \\
 HE0336+0113    &  5700 & 3.5 & 1.6 &  12,200 & 100 &  Magellan \\
 HE1031$-$0020  &  5080 & 2.2 & 1.6 & 2400 & 80 & P200 \\
 HE1150$-$0428  &  5200 & 2.55 & 1.6 &  2400 & 70 & P200\\ 
 HE1410$-$0004  &  5605 & 3.5 &  1.6 & 2400\tablenotemark{b} & 60 & P200 \\
 HE1410+0213    & 4985 & 2.0 & 1.8 & 2700 & 80 & P200 \\
 HE1434$-$1442  & 5420 & 3.15 & 1.6 & 6000\tablenotemark{c} & 73 & P200 \\
 HE1443+0113    &  4945 & 1.95 & 1.8 & 550\tablenotemark{c,e} & low & P200 \\
 HE1509$-$0806  & 5185 & 2.5 & 1.6 & 2400 & 70 & P200 \\     
 HE2158$-$0348  & 5215 & 2.5 & 1.6 &  14,400 & 100 &  Magellan \\
 HE2232$-$0603  &  5750 & 3.5 & 1.6 & 18,000 & 90 &    Magellan  \\
 HE2356$-$0410  &  5205 & 2.5 & 1.8 & 6000 & 100 &  Magellan \\
\enddata
\tablenotetext{a}{Signal to noise ratio in the continuum near
4570~\AA\ per 4 pixel spectral resolution element.}
\tablenotetext{b}{HIRES-R spectrum with new mosaic CCD   detector.}
\tablenotetext{c}{HIRES-B spectrum with new mosaic CCD   detector.}
\tablenotetext{d}{Assumes second component does not affect the observed colors.}
\tablenotetext{e}{Only one short exposure available, stopped by fog.}
\end{deluxetable}

\clearpage

%
% Table 3a
%
\clearpage

\begin{deluxetable}{lllr rrrrr}
\tablenum{3}
\tablecaption{Equivalent Widths for the First 5 Stars of for the Primary Sample of 
C-Stars from the HES \label{table_eqwa}}
\tablewidth{0pt}
\tablecolumns{10}
\tablehead{\colhead{Line $\lambda$} &
\colhead{Species} &
\colhead{EP} &
\colhead{Log($gf$)} &
\colhead{HE0012} &
\colhead{HE0058} &
\colhead{HE0143} &
\colhead{HE0212} &
\colhead{HE0336}  \\
\colhead{(\AA )} & \colhead{} & \colhead{(eV)} & \colhead{(dex)} &
\colhead{$-$1441\tablenotemark{a}} &
\colhead{$-$0244} &
\colhead{$-$0441} & 
\colhead{$-$0557} &
\colhead{+0113}  \\
\colhead{} & \colhead{} & \colhead{} & \colhead{} &
\colhead{(m\AA )} & \colhead{(m\AA )} & 
\colhead{(m\AA )} & \colhead{(m\AA )} & \colhead{(m\AA )} 
}
\startdata
 4057.52 & Mg~I  &   4.34 &  $-$1.200 &    34.0 &    19.2 &    21.2 &  \nodata  &    34.4  \\
 4703.00 & Mg~I  &   4.34 &  $-$0.670 &   100.7 &  \nodata  &  \nodata  &  \nodata  &  \nodata   \\
 5172.70 & Mg~I  &   2.71 &  $-$0.380 &   258.9 &   144.7 &   163.5 &   183.5 &   210.9  \\
 5183.62 & Mg~I  &   2.72 &  $-$0.160 &   357.5 &   172.1 &   187.8 &  \nodata  &   340.9  \\
 3961.52 & Al~I  &   0.00 &  $-$0.340 &  \nodata  &   113.2 &    77.8 &   150.0 &   117.1  \\
 3905.53 & Si~I  &   1.91 &  $-$1.040 &  \nodata  &  \nodata  &   239.1 &  \nodata  &  \nodata   \\
 4226.74 & Ca~I  &   0.00 &   0.240 &   303.0 &  \nodata  &  \nodata  &  \nodata  &  \nodata   \\
 4425.44 & Ca~I  &   1.88 &  $-$0.360 &  \nodata  &  \nodata  &    25.6 &  \nodata  &  \nodata   \\
\enddata
\tablenotetext{a}{\eqw\ given as a guide.  Synthesis used throughout
for this binary star.  See text.}
\tablenotetext{b}{Synthesis used to derive Pb abundance. \eqw\ given as a guidance
to line strength.}
\end{deluxetable}

\begin{deluxetable}{lllr rrrrr}
\tablenum{4}
\tablecaption{Equivalent Widths for the Next 5 Stars of for the Primary Sample of 
C-Stars from the HES \label{table_eqwb}}
\tablewidth{0pt}
\tablecolumns{10}
\tablehead{\colhead{Line $\lambda$} &
\colhead{Species} &
\colhead{EP} &
\colhead{Log($gf$)} &
\colhead{HE1031} &
\colhead{HE1150} &
\colhead{HE1410} &
\colhead{HE1410} &
\colhead{HE1434}  \\
\colhead{(\AA )} & \colhead{} & \colhead{(eV)} & \colhead{(dex)} &
\colhead{$-$0020} &
\colhead{$-$0428} &
\colhead{+0213} & 
\colhead{$-$0004} &
\colhead{$-$1442} 
  \\
\colhead{} & \colhead{} & \colhead{} & \colhead{} &
\colhead{(m\AA )} & \colhead{(m\AA )} & 
\colhead{(m\AA )} & \colhead{(m\AA )} & \colhead{(m\AA )} 
}
\startdata
 4057.52 & Mg~I  &   4.34 &  $-$1.200 &    18.0 &  \nodata  &  \nodata  &    13.9 &  \nodata   \\
 4703.00 & Mg~I  &   4.34 &  $-$0.670 &  \nodata  &  \nodata  &  \nodata  &    31.4 &  \nodata   \\
 5172.70 & Mg~I  &   2.71 &  $-$0.380 &   160.3 &   115.2 &   218.6 &   128.6 &   191.7  \\
 5183.62 & Mg~I  &   2.72 &  $-$0.160 &   189.2 &   130.4 &   246.3 &   142.9 &   241.6  \\
 3961.52 & Al~I  &   0.00 &  $-$0.340 &   174.5 & \nodata &  \nodata  &  \nodata  &  \nodata   \\
 4435.69 & Ca~I  &   1.89 &  $-$0.520 &    55.0 &    32.4 &  \nodata  &  \nodata  &  \nodata   \\
 4454.79 & Ca~I  &   1.90 &   0.260 &   110.0 &    74.0 &  \nodata  &  \nodata  &  \nodata   \\
 4578.56 & Ca~I  &   2.52 &  $-$0.558 &    36.0 &     9.0 &  \nodata  &  \nodata  &  \nodata   \\
\enddata
%\tablenotetext{a}{Upper limits to \eqw\ are indicated by negative numbers.}
\tablenotetext{a}{Synthesis used to derive Pb abundance. \eqw\ given as a guidance
to line strength.}
\end{deluxetable}

\begin{deluxetable}{lllr rrrrr}
\tablenum{5}
\tablecaption{Equivalent Widths for the Last 5 Stars of for the Primary Sample of 
C-Stars from the HES \label{table_eqwc}}
\tablewidth{0pt}
\tablecolumns{10}
\tablehead{\colhead{Line $\lambda$} &
\colhead{Species} &
\colhead{EP} &
\colhead{Log($gf$)} &
\colhead{HE1443} &
\colhead{HE1509} &
\colhead{HE2158} &
\colhead{HE2232} &
\colhead{HE2356}  \\
\colhead{(\AA )} & \colhead{} & \colhead{(eV)} & \colhead{(dex)} &
\colhead{+0113} &
\colhead{$-$0806} &
\colhead{$-$0348} & 
\colhead{$-$0603} &
\colhead{$-$0410}  \\
\colhead{} & \colhead{} & \colhead{} & \colhead{} &
\colhead{(m\AA )} & \colhead{(m\AA )} & 
\colhead{(m\AA )} & \colhead{(m\AA )} & \colhead{(m\AA )} 
}
\startdata
 4057.52 & Mg~I  &   4.34 &  $-$1.200 &  \nodata &  \nodata &    29.0 &    72.0 &    18.0  \\
 4167.28 & Mg~I  &   4.34 &  $-$1.000 &  \nodata &  \nodata &  \nodata &   100.0 &  \nodata  \\
 4703.00 & Mg~I  &   4.34 &  $-$0.670 &  \nodata &  \nodata &  \nodata &   105.4 &  \nodata  \\
 5172.70 & Mg~I  &   2.71 &  $-$0.380 &   251.3 &   167.0 &   187.1 &   275.9 &   123.2  \\
 5183.62 & Mg~I  &   2.72 &  $-$0.160 &   333.8 &   211.0 &   237.5 &   356.6 &   141.9  \\
 3961.52 & Al~I  &   0.00 &  $-$0.340 &  \nodata &   119.0 &   129.1 &   140.0 &   111.3  \\
 4435.69 & Ca~I  &   1.89 &  $-$0.520 &  \nodata &    29.0 &    56.0 &    60.3 &    26.7  \\
 4454.79 & Ca~I  &   1.90 &   0.260 &  \nodata &    74.0 &    94.0 &    95.1 &    79.6  \\
\enddata
%\tablenotetext{a}{Upper limits to \eqw\ are indicated by negative numbers.}
\tablenotetext{a}{Synthesis used to derive Pb abundance. \eqw\ given as a guidance
to line strength.}
\end{deluxetable}

\clearpage

%
% Table 3d
%
\clearpage
\begin{deluxetable}{lllr rrr}
\tablenum{6}
\tablecaption{Equivalent Widths for Redder Lines in the Spectra of Three 
C-Stars \label{table_eqwd}}
\tablewidth{0pt}
\tablecolumns{10}
\tablehead{\colhead{Line $\lambda$} &
\colhead{Species} &
\colhead{EP} & \colhead{Log($gf$)} & 
\colhead{HE1410$-$0004} &
\colhead{HE1434$-$1442} &
\colhead{HE1443+0113} \\
\colhead{(\AA )} & \colhead{} & \colhead{(eV)} & \colhead{(dex)} &
\colhead{($W_{\lambda}$ -- m\AA )} & \colhead{($W_{\lambda}$ -- m\AA )}
& \colhead{($W_{\lambda}$ -- m\AA )}
}
\startdata
  6707.76 & Li~I &   0.00 & 0.178 &  \llap{$<$}10.0 & \nodata & \nodata \\
  6300.30  & O~I &    0.00 &  $-9.78$  &  \llap{$<$}6.0
           & \nodata & \nodata \\
  7771.94   & O~I &   9.15  &  0.369 & 7.0 & \nodata & \nodata \\
%  6300 upper limit in HE1410-0004, 7771 is a probable detection  
  5688.19  & Na~I &    2.10 &  $-0.420$ & 7.0 & \nodata & \nodata \\
  5889.95 &  Na~I &   0.00 &   0.110  & 148.5  &  178.8 & 363.0 \\
  5895.92 &  Na~I &    0.00 & $-$0.190 &  127.0 &  176.5 & 210.0 \\
  5528.41  & Mg~I &    4.34 &  $-0.480$ & 29.5 & 71.8 & \nodata \\
  5711.09 & Mg~I &   4.34 &  $-0.167$ & 6.0 & \nodata & \nodata \\
\enddata
%\tablenotetext{a}{Upper limits to \eqw\ are indicated by negative numbers.}
\end{deluxetable}      

% file /scr2/jlc/hamburg_survey/paper_cstars_2005/table3d_26sep2005.tex

\clearpage

\clearpage   
\begin{deluxetable}{l rrr}
\tablenum{7}
\tablewidth{0pt}
\small
\tablecaption{Fit Fe~I Slopes With EP, Equivalent Width and Wavelength 
\label{table_slopes}}
\tablehead{\colhead{Star ID} &
\colhead{$\Delta$[X/Fe]/$\Delta$(EP)\tablenotemark{a}} &
\colhead{$\Delta$[X/Fe]/${\Delta}[W_{\lambda}/\lambda]$} &
\colhead{$\Delta$[X/Fe]/${\Delta}\lambda$} \\ 
\colhead{} & \colhead{(dex/eV)} & \colhead{(dex)} & 
\colhead{($10^{-4}$dex/$\AA$)}
}
\startdata
HE0012$-$1441 &  \tablenotemark{b} & \tablenotemark{b} & \tablenotemark{b} \\
HE0058$-$0244 & $-$0.013 & $-$0.198  & 1.02 \\
HE0143$-$0441 &  $-$0.026 &  0.024 & 0.65 \\
HE0212$-$0557 & 0.088  &  0.207 & $-$5.33 \\
HE0336+0113 & 0.075 & $-0.121$ & 0.98 \\
HE1031$-$0020 &  0.039 &  $-0.117$ & 2.02 \\
HE1150$-$0428 &   $-0.066$ &  $-0.027$ & 0.14 \\
HE1410$-$0004 &  $-0.014$ & 0.001 & $-0.08$ \\
HE1410+0213 & \tablenotemark{c} & \tablenotemark{c} & \tablenotemark{c} \\
HE1434$-$1442 & 0.056 &  0.006 & $-$1.38 \\
HE1443+0113 & \tablenotemark{d} & \tablenotemark{d} & \tablenotemark{d} \\
HE1509$-$0806 & \tablenotemark{e} & \tablenotemark{e} & \tablenotemark{e} \\
HE2158$-$0348 & 0.068  & $-0.248$ &  1.63 \\
HE2232$-$0603 & 0.057 & $-0.013$ &  1.87 \\
HE2356$-$0410 & $-0.036$ &  0.006 &  $-$0.83 \\
\enddata
\tablenotetext{a}{Typical range of EP is 3 eV.}
\tablenotetext{b}{There were only 14 measured Fe~I lines in this star.}
\tablenotetext{c}{There were only 12 detected Fe~I lines in this star.}
\tablenotetext{d}{There were only 10 measured Fe~I lines in this star.}
\tablenotetext{e}{There were only 17 detected Fe~I lines in this star.}
\end{deluxetable}

% HE0012-1441    
% HE0058-0244    
% HE0143-0441    
% HE0212-0557    
\begin{deluxetable}{l  rrrr rrrr rrrr rrrr}
\tablenum{8}
% \tiny
\tabletypesize{\tiny}
\rotate
\tablewidth{0pt}
\tablecaption{Abundances for the First Four EMP C-Stars From the HES \label{table_abunda}}
\tablehead{\colhead{Species} & 
\multispan{4}{HE0012$-$1441} &
\multispan{4}{HE0058$-$0244} &
\multispan{4}{HE0143$-$0441} &
\multispan{4}{HE0212$-$0557} \\
\colhead{} & 
\colhead{[X/Fe]} & \colhead{log$\epsilon(X)$} & \colhead{No.} & \colhead{$\sigma$} &  
\colhead{[X/Fe]} & \colhead{log$\epsilon(X)$} & \colhead{No.} & \colhead{$\sigma$} &
\colhead{[X/Fe]} & \colhead{log$\epsilon(X)$} & \colhead{No.} & \colhead{$\sigma$} &
\colhead{[X/Fe]} & \colhead{log$\epsilon(X)$} & \colhead{No.} & \colhead{$\sigma$} \\
\colhead{} & \colhead{(dex)} & \colhead{(dex)} & \colhead{Lines} & \colhead{(dex)} &
\colhead{(dex)} & \colhead{(dex)} & \colhead{Lines} & \colhead{(dex)} &
\colhead{(dex)} & \colhead{(dex)} & \colhead{Lines} & \colhead{(dex)} &
\colhead{(dex)} & \colhead{(dex)} & \colhead{Lines} & \colhead{(dex)} 
}
\startdata 
C$-$CH  &   1.59 &   7.66   &    1 &  \nodata   &   1.92 &   7.76   &    1 &  \nodata   &  
     1.98 &   8.26   &    1 &  \nodata   &   1.74 &   8.06   &    1 &  \nodata   \\
N$-$CN  &   0.64 &   6.05   &    1 &  \nodata   &   1.77 &   6.95   &    1 &  \nodata   &   
   1.73\tablenotemark{d} &   7.35\tablenotemark{d}   &    1 &  \nodata   &   1.09 &   6.75   &    1 &  \nodata   \\
MgI   &   0.91 &   5.93   &    4 &    0.16   &   0.54 &   5.33   &    3 &    0.19   &   0.63 &   5.86   &    3 &    0.08   &   0.04 &   5.32   &    1 &  \nodata   \\
AlI   & \nodata & \nodata   &  \nodata &  \nodata    &   0.34 &   4.06   &    1 &  \nodata   &  $-$0.22 &   3.94   &    1 &  \nodata   &   0.01 &   4.21   &    1 &  \nodata   \\
CaI   &   0.42 &  4.26  &    3 &    0.11   &   0.96 &   4.57   &    2 &    0.20   &   0.43 &   4.48   &    2 &    0.17   &   0.14 &   4.23   &    1 &  \nodata   \\
ScII  & \nodata & \nodata   &  \nodata &  \nodata    &   0.56 &   0.91   &    1 &  \nodata   &   0.67 &   1.46   &    1 &  \nodata   & \nodata & \nodata   &  \nodata &  \nodata    \\
TiI   &   0.03 &   2.50   &    2 &    0.10   &   0.57 &   2.81   &   10 &    0.17   &   0.54 &   3.22   &    5 &    0.26   &   0.14 &   2.87   &    5 &    0.32   \\
TiII  &   0.13 &   2.60   &    5 &    0.08   &   0.55 &   2.79   &   10 &    0.16   &   0.26 &   2.94   &    8 &    0.16   &   0.32 &   3.04   &    8 &    0.32   \\
\enddata
\tablenotetext{a}{This is [Fe/H].}
\tablenotetext{b}{Pb~I 4057 line is swamped by molecular features.}
\tablenotetext{c}{No Pb abundance as the large negative $v_r$ moves 
the 4057~\AA\ Pb~I line onto a CCD defect.}
\tablenotetext{d}{This value supercedes that given in \cite{cohen04}.}
\end{deluxetable}

% HE0336+0113   
% HE1031-0020    
% HE1150-0428
% HE1410+0213    
\begin{deluxetable}{l  rrrr rrrr rrrr rrrr}
\tablenum{9}
% \tiny
\tabletypesize{\tiny}
\rotate
\tablewidth{0pt}
\tablecaption{Abundances for the Second Four EMP C-Stars From the HES \label{table_abundb}}
\tablehead{\colhead{Species} & 
\multispan{4}{HE0336+0113} &
\multispan{4}{HE1031$-$0020} &
\multispan{4}{HE1150$-$0428} &
\multispan{4}{HE1410+0213}  \\
\colhead{} & 
\colhead{[X/Fe]} & \colhead{log$\epsilon(X)$} & \colhead{No.} & \colhead{$\sigma$} &
\colhead{[X/Fe]} & \colhead{log$\epsilon(X)$} & \colhead{No.} & \colhead{$\sigma$} &
\colhead{[X/Fe]} & \colhead{log$\epsilon(X)$} & \colhead{No.} & \colhead{$\sigma$} &
\colhead{[X/Fe]} & \colhead{log$\epsilon(X)$} & \colhead{No.} & \colhead{$\sigma$} \\
\colhead{} & \colhead{(dex)} & \colhead{(dex)} & \colhead{Lines} & \colhead{(dex)} &
\colhead{(dex)} & \colhead{(dex)} & \colhead{Lines} & \colhead{(dex)} &
\colhead{(dex)} & \colhead{(dex)} & \colhead{Lines} & \colhead{(dex)} &
\colhead{(dex)} & \colhead{(dex)} & \colhead{Lines} & \colhead{(dex)} 
}
\startdata 
C$-$CH & 2.25 & 8.16 & 1 & \nodata & 1.63 & 7.36 & 1 & \nodata &
     2.37 & 7.66 & 1 & \nodata & 1.73 & 8.16 & 1 & \nodata \\
N$-$CN & 1.60 & 6.85 & 1 & \nodata & 2.48 & 7.55 & 1 & \nodata &
     2.52 & 7.15 & 1 & \nodata & 1.78 & 7.55 & 1 & \nodata \\
NaI   & \nodata & \nodata   &  \nodata &  \nodata    & \nodata & \nodata   &  \nodata
 &  \nodata    & \nodata & \nodata   &  \nodata &  \nodata    & \nodata & \nodata   &  \nodata &  \nodata    \\
MgI   &   1.04 &   5.90   &    3 &    0.18   &   0.50 &   5.19   &    3 &    0.18   &  
     0.34 &   4.58   &    2 &    0.06   &   0.18 &   5.56   &    2 &    0.03   \\
AlI   &   0.34 &   4.13   &    1 &  \nodata   &   0.88 &   4.49   &    1 &  \nodata   & 
   \nodata   & \nodata & \nodata   &  \nodata & \nodata   & \nodata & \nodata   &  \nodata   \\
SiI   & \nodata & \nodata   &  \nodata &  \nodata    & \nodata & \nodata   &  \nodata 
 &  \nodata    & \nodata & \nodata   &  \nodata &  \nodata    & \nodata & \nodata   &  \nodata &  \nodata    \\
CaI   &   0.42 &   4.10   &    1 &  \nodata   &   1.12 &   4.62   &    3 &    0.22  
  &   1.03 &   4.09   &    3 &    0.04   & \nodata & \nodata   &  \nodata &  \nodata    \\
TiI   &   0.39 &   2.70   &    5 &    0.19   &   0.46 &   2.60   &   11 &    0.31  
  &   0.49 &   2.17   &    5 &    0.12   &   0.29 &   3.12   &    2 &    0.04   \\
\enddata 
\tablenotetext{a}{This is [Fe/H].}  
\tablenotetext{b}{Pb~I 4057 line is swamped by molecular features.}
\tablenotetext{c}{No Pb abundance as the large negative $v_r$ moves 
the 4057~\AA\ Pb~I line onto a CCD defect.}
\end{deluxetable}

% HE1410-0004    
% HE1434-1442 
% HE1443+0113    
% HE1509-0806    
\begin{deluxetable}{l  rrrr rrrr rrrr rrrr}
\tablenum{10}
% \tiny
\tabletypesize{\tiny}
\rotate
\tablewidth{0pt}
\tablecaption{Abundances for the Third Four EMP C-Stars From the HES \label{table_abundc}}
\tablehead{\colhead{Species} &
\multispan{4}{HE1410$-$0004}  & 
\multispan{4}{HE1434$-$1442}  & 
\multispan{4}{HE1443+0113} &
\multispan{4}{HE1509$-$0806} \\
\colhead{} & 
\colhead{[X/Fe]} & \colhead{log$\epsilon(X)$} & \colhead{No.} & \colhead{$\sigma$} &
\colhead{[X/Fe]} & \colhead{log$\epsilon(X)$} & \colhead{No.} & \colhead{$\sigma$} &
\colhead{[X/Fe]} & \colhead{log$\epsilon(X)$} & \colhead{No.} & \colhead{$\sigma$} &
\colhead{[X/Fe]} & \colhead{log$\epsilon(X)$} & \colhead{No.} & \colhead{$\sigma$} \\
\colhead{} & \colhead{(dex)} & \colhead{(dex)} & \colhead{Lines} & \colhead{(dex)} &
\colhead{(dex)} & \colhead{(dex)} & \colhead{Lines} & \colhead{(dex)} &
\colhead{(dex)} & \colhead{(dex)} & \colhead{Lines} & \colhead{(dex)} &
\colhead{(dex)} & \colhead{(dex)} & \colhead{Lines} & \colhead{(dex)} 
}
\startdata  
Li~I   &   $<$1.03\tablenotemark{d} & $<$1.32 & 1 & \nodata & \nodata &  \nodata    & \nodata & \nodata   &
    \nodata &  \nodata    & \nodata & \nodata   &    \nodata &  \nodata    & \nodata & \nodata \\
C$-$CH  &   1.99 &   7.56   &    1 &  \nodata   &   1.95 &   8.16   &    1 &  \nodata   &   1.84 &   8.36   &    1 &  \nodata   &   1.98 &   7.66   &    1 &  \nodata   \\
N$-$CN  & \nodata & \nodata   &  \nodata &  \nodata    &   1.40 &   6.95   &    1 &  \nodata   & \nodata & \nodata   &  \nodata &  \nodata    &   2.23 &   7.25   &    1 &  \nodata   \\
O~I     &  1.18 &   6.90   &    1 &  \nodata   & \nodata & \nodata   &  \nodata &  \nodata    & \nodata & \nodata   &  \nodata &  \nodata    & \nodata & \nodata   &  \nodata &  \nodata    \\
NaI   &   0.48 &   3.78   &    3 &    0.20   &   0.03 &  3.97   &    4 &    0.23   & 
   0.37 &   4.62   &    2 &    0.27   & \nodata & \nodata   &  \nodata &  \nodata    \\
MgI   &   0.58 &   5.10   &    6 &    0.29   &   0.30 &   5.45   &    3 &    0.05   &   0.37 &   5.84   &    2 &    0.09   &   0.64 &   5.27   &    2 &    0.11   \\
AlI   &   \nodata & \nodata   &  \nodata &  \nodata   & \nodata & \nodata   &  \nodata &  \nodata  &
     \nodata & \nodata   &  \nodata &  \nodata    &   0.28 &   3.84   &    1 &  \nodata   \\
SiI   &  \nodata & \nodata   &  \nodata  &  \nodata   &   0.74 &   5.91   &    2 &    0.26   & \nodata & \nodata   &  \nodata &  \nodata    & \nodata & \nodata   &  \nodata &  \nodata    \\
\enddata
\tablenotetext{a}{This is [Fe/H].}
\tablenotetext{b}{Pb~I 4057 line is swamped by molecular features.}
\tablenotetext{c}{No Pb abundance as the large negative $v_r$ moves 
the 4057~\AA\ Pb~I line onto a CCD defect.}
\tablenotetext{d}{The meteoritic abundance is used for Li, rather than the solar one.}
\end{deluxetable}

% HE2158$-$0348    
% HE2232$-$0603    
% HE2356$-$0410
\begin{deluxetable}{l  rrrr rrrr rrrr}
\tablenum{11}
% \tiny
\tabletypesize{\tiny}
%\rotate
\tablewidth{0pt}
\tablecaption{Abundances for the Last Three EMP C-Stars From the HES \label{table_abundd}}
\tablehead{\colhead{Species} &
\multispan{4}{HE2158$-$0348}  & 
\multispan{4}{HE2232$-$0603} &
\multispan{4}{HE2356$-$0410} \\
\colhead{} & 
\colhead{[X/Fe]} & \colhead{log$\epsilon(X)$} & \colhead{No.} & \colhead{$\sigma$} &
\colhead{[X/Fe]} & \colhead{log$\epsilon(X)$} & \colhead{No.} & \colhead{$\sigma$} &
\colhead{[X/Fe]} & \colhead{log$\epsilon(X)$} & \colhead{No.} & \colhead{$\sigma$} \\
\colhead{} & \colhead{(dex)} & \colhead{(dex)} & \colhead{Lines} & \colhead{(dex)} &
\colhead{(dex)} & \colhead{(dex)} & \colhead{Lines} & \colhead{(dex)} &
\colhead{(dex)} & \colhead{(dex)} & \colhead{Lines} & \colhead{(dex)} 
}
\startdata 
C$-$CH  &   1.87 &   7.76   &    1 &  \nodata   &   1.22 &   7.96   &    1 &  \nodata   &   2.14 &   7.66   &    1 &  \nodata \\
N$-$CN  &   1.52 &   6.75   &    1 &  \nodata   &   0.47 &   6.55   &    1 &  \nodata   &   1.89 &   6.75   &    1 &  \nodata \\
MgI   &   0.68 &   5.52   &    3 &    0.18   &   0.85 &   6.54   &    5 &    0.32   &   0.36 &   4.83   &    3 &    0.50  \\
AlI   &   0.47 &   4.24   &    1 &  \nodata   &   0.26 &   4.88   &    1 &  \nodata   &   0.25 &   3.65   &    1 &  \nodata \\
CaI   &   0.82 &   4.48   &    2 &    0.05   &   0.35 &   4.86   &    2 &    0.15   &   0.71 &   4.01   &    2 &    0.13  \\
ScII  & \nodata & \nodata   &  \nodata &  \nodata    &   0.05 &   1.31   &  1 &  \nodata   &   0.48 &   0.51   &    1 &  \nodata \\
TiI   &   0.43 &   2.72   &    6 &    0.24   &   0.35 &   3.50   &   13 &    0.21   &   0.25 &   2.17   &    6 &    0.13 \\
TiII  &   0.65 &   2.94   &    8 &    0.28   &   0.11 &   3.25   &   11 &    0.17   &   0.33 &   2.25   &    8 &    0.23 \\
\enddata
\tablenotetext{a}{This is [Fe/H].}
\end{deluxetable}

\clearpage

\begin{deluxetable}{lcc}
\tablenum{12}
% \tiny
% \tabletypesize{\tiny}
\tablewidth{0pt}
\tablecaption{$^{12}$C/$^{13}$C Ratios for EMP C-Stars From the HES
\label{table_c12c13}}
\tablehead{\colhead{ID} & \colhead{$^{12}$C/$^{13}$C(C$_2$)\tablenotemark{a}} &
\colhead{$^{12}$C/$^{13}$C(CH)}
}
\startdata
HE0007$-$1832 & \llap{$>$}2.0 & \nodata \\
HE0012$-$1441 & \llap{$>$}3.0 & \nodata \\
HE0058$-$0244 & 3.5 & 8 -- 10 \\
HE0143$-$0411 & \llap{$>$}4.0 & \nodata \\
HE0212$-$0557 & 4.0 & 3 -- 4 \\
HE0336+0113 & 2.5 & 7.5 \\
HE1031$-$0020 & 5.0 & \nodata \\
HE1150$-$0428 & 4.0 & \nodata \\
HE1410+0213 & 2.0 & 2.5 \\
HE1410$-$0004 & \llap{$>$}3.0 & \nodata \\
HE1434$-$1442 & 5.0 & \nodata \\
HE1443+0113 & 5.0 & \nodata \\
HE1509$-$0806 & 4.0 & \nodata \\
HE2158$-$0348 & 6.0 & 3 -- 5 \\
HE2232$-$0603 & \llap{$>$}6.0 & $\ge$30 \\
HE2356$-$0410 & 4.0 & 3 -- 5 \\
\enddata
\tablenotetext{a}{The uncertainty in the deduced \ciso\ ratios is 30\% of the
isotopic ratio.}
\end{deluxetable}

\clearpage

\begin{deluxetable}{l rrrrr r}
\tablenum{13}
\tablewidth{0pt}
\tabletypesize{\small}
\tablecaption{Abundance Changes for Small Changes in Stellar 
Parameters\tablenotemark{a}
\label{table_sens}}
\tablehead{\colhead{Species} & \colhead{${\Delta}$[X/Fe]} &
\colhead{$\Delta$[X/Fe]} &
\colhead{$\Delta$[X/Fe]} &  \colhead{$\Delta$[X/Fe]} &  
\colhead{$\Delta$[X/Fe]} & 
\colhead{$\Delta$[X/Fe]\tablenotemark{b}} \\
\colhead{} & \colhead{(\teff$-150$K)} & \colhead{(\grav$-0.4$ dex)} &
\colhead{Model[Fe/H]-0.5} & \colhead{($v_t$-0.2~\kms)} & 
\colhead{\eqw\ Unc.} &
\colhead{1 Star} \\
\colhead{} & \colhead{(dex)} & \colhead{(dex)} & \colhead{(dex)} &
\colhead{(dex)} & \colhead{(dex)} & \colhead{(1$\sigma$)(dex)} 
}
\startdata
C(CH) & $-0.18$ & 0.14 & $-0.04$ & $-0.02$ & \nodata & 0.23 \\
N(CN)\tablenotemark{f} & 0.04 & 0.09 & 0.00 & 0.00 & \nodata & 0.26\tablenotemark{g} \\
MgI & 0.00 &  0.11 &  0.01 &  0.03 &  0.05 &    0.13  \\
AlI & $-0.02$ &  0.10 &  0.02 &  0.10 &  0.08 &    0.16 \\
CaI & 0.04 &  0.03 &  0.01 &  0.06 &  0.06 &    0.10 \\
TiI & $-0.01$ & $-0.01$ & $-0.01$ &  0.02 &  0.03 &    0.04 \\
TiII & 0.08 & $-0.14$ & $-0.01$ &  0.11 &  0.03 &    0.20 \\
TiII\tablenotemark{h} & $-0.06$ &  0.03 &  0.01 &  0.11 &  0.03 &    0.13 \\
\enddata
\tablenotetext{a}{Computed from the line list and the stellar
parameters of HE2158$-$0348 with respect to Fe~I for all species.}
\tablenotetext{b}{1$\sigma$ uncertainty in [X/Fe] for a single (typical) star
including the 5 sources of uncertainties.}
\tablenotetext{c}{Treated as individual absorption lines without HFS corrections.}
\tablenotetext{d}{Change in log$[\epsilon$(Fe)].}
\tablenotetext{e}{4057~\AA\ feature treated as a single Pb~I line.}
\tablenotetext{f}{Assumes log$\epsilon$(C) varies as for CH.}
\tablenotetext{g}{This includes the uncertainty in $\epsilon$(C), as $\epsilon$(N)
is derived from lines of CN.}
\tablenotetext{h}{Changes computed with respect to Fe~II.}
\end{deluxetable}

\clearpage

\begin{deluxetable}{l rrrrr}
\tablenum{14}
\tablewidth{0pt}
\small
\tablecaption{Additional C-Stars From the Literature 
\label{table_add}}
\tablehead{\colhead{Star ID} & \colhead{Ref.\tablenotemark{a}} &
\colhead{Binary} &
\colhead{Period, $v_r$ Amp} & \colhead{[Ba/Fe]} & \colhead{$^{12}$C/$^{13}$C} \\ 
\colhead{} & \colhead{} & \colhead{} & \colhead{(days, \kms)} &
\colhead{(dex)} 
}
\startdata
G77--61 & 1,2 & Yes &  245, 20  & \nodata & \nodata \\
CS22880--074 & 8 & \nodata & \nodata & +1.34 & \llap{$>$}40 \\
CS22881--036 & 8 & \nodata & \nodata & +1.93 & 40\rlap{:}  \\
CS22898--027 & 8 & \nodata & \nodata & +2.27 & \llap{$>$}20 \\
CS22942-019 & 4, 8 & Yes & 2800, 5 &  +1.92 & 30\rlap{:} \\
CS22948--027 & 8,9 & Yes & 505, 5 & +1.67 & 14 \\
CS29497--030 & 6,7 & Yes & 342, 4 & +2.17 & \llap{$>$}10 \\
CS29497--034 & 9 & \nodata & \nodata & $+$2.03 & 12 \\
CS29498--043 & 3 & \nodata & \nodata & $-$0.45 & 6 \\
CS30301--015 & 4 & \nodata & \nodata & $+$1.45 & \nodata \\
CS31062--050 & 4,5 & \nodata & \nodata & $+$2.61 & \nodata \\
\enddata
\tablenotetext{a}{1. \cite{plez05}, \cite{plez06} 2. \cite{dearborn},
3. \cite{aoki02b}, 4. \cite{aoki02c},
5. \cite{johnson04},
6. \cite{sivarani}, 7. \cite{preston00}, 8. \cite{preston01},
9. \cite{hill00}. } 
\end{deluxetable}

\clearpage   

\begin{deluxetable}{l r rrrr c}
\tablenum{15}
\tablewidth{0pt}
\small
\tablecaption{Statistics for Selected Abundance Ratios For the Primary
Sample of \ncstars\ C-Stars From
the HES\tablenotemark{a} \label{table_ratios}}
\tablehead{\colhead{Species\tablenotemark{h}} & \colhead{Nstars\tablenotemark{b}} &
\colhead{Min. [X/Y]} &
\colhead{Max. [X/Y]} & 
\colhead{Median [X/Y]} &
\colhead{$\sigma$[X/Y]} & \colhead{EMP Dwarfs\tablenotemark{c}} \\
\colhead{} & \colhead{} &
\colhead{(dex)} & \colhead{(dex)} & \colhead{(dex)} &
\colhead{(dex)} & \colhead{(dex)} 
}
\startdata
$[$C/Fe$]$    & 16 &    1.22 & 2.52  &  1.93  & 0.31  & 0.2\tablenotemark{f} \\
$[$N/Fe$]$    & 14 &    0.47 & 2.52  &   1.75  & 0.59   &  0.0\tablenotemark{f} \\
$[$Na/Fe$]$    & 3 &    0.03 & 0.48  &  0.37  & 0.23   &  0.41\tablenotemark{g} \\
$[$Mg/Fe$]$   & 12 &    0.04 & 1.04  & 0.55    & 0.27  &  0.56 \\
$[$Al/Fe$]$\tablenotemark{d}
     & 10 &  $-0.55$ & 0.88  &  0.27  & 0.39 & \llap{$-$}0.09 \\
$[$Ca/Fe$]$   & 14 &   0.11 & 1.12  &   0.42 & 0.34  &  0.31 \\
$[$ScII/Fe$]$\tablenotemark{d}
    &  5 &  0.05 &  0.67  & 0.48  & 0.26 &    0.24  \\
$[$TiI/Fe$]$  & 14  &  0.03 & 0.57 &   0.36  & 0.15  &   0.36 \\
\enddata
\tablenotetext{a}{We include the dwarf C-star HE0007$-$1832 
from \cite{cohen04}.}
\tablenotetext{b}{Upper limits are ignored.}
\tablenotetext{c}{Data from \cite{cohen04}.}
\tablenotetext{d}{Only 1 line used, with possible blending
by molecular features.}
\tablenotetext{e}{Here we exclude the four C-stars with [Ba/Fe]$< 0.2$ dex.}
\tablenotetext{f}{Data for unmixed giants from \cite{spite05}.}
\tablenotetext{g}{Data for giants from \cite{cayrel03}.}
\tablenotetext{h}{Detection in a minimum of three stars is required.}
\end{deluxetable}

\clearpage

% Minimum is detection in 3 stars

\clearpage

\clearpage

\end{document}